\documentclass[12pt]{iopart}

\usepackage{iopams}
\usepackage{graphicx}
\usepackage[normal]{caption}
\usepackage{color}

\begin{document}

\title[Computation of many-particle quantum trajectories with exchange interaction]{Computation of many-particle quantum trajectories with exchange interaction:
Application to the simulation of nanoelectronic devices}

\author{A. Alarc\'{o}n, S. Yaro, X. Cartoix\`{a} and X. Oriols}

\address{Departament d\rq{}Enginyeria Electr\`{o}nica, Universitat Aut\`{o}noma de Barcelona, 08193, Bellaterra, SPAIN}
\ead{xavier.oriols@uab.es}

\begin{abstract}
Following Ref. [Oriols X 2007 Phys. Rev. Lett., {\bf 98} 066803], an algorithm to deal with the exchange interaction in non-separable quantum systems is presented. The algorithm can be applied to fermions or bosons and, by construction, it exactly ensures that any observable is totally independent from the interchange of particles. It is based on the use of conditional Bohmian wave functions which are solutions of  single-particle pseudo-Schr\"{o}dinger equations. The exchange symmetry is directly defined by demanding symmetry properties of the quantum trajectories in the configuration space with a universal algorithm, rather than through a particular exchange-correlation functional introduced into the single-particle pseudo-Schr\"{o}dinger equation. It requires the computation of $N^2$ conditional wave functions to deal  with $N$ identical particles. For separable Hamiltonians, the algorithm reduces to the standard Slater determinant for fermions, or permanent for bosons. A numerical test for a two-particle system, where exact solutions for non-separable Hamiltonians are computationally accessible, is presented. The numerical viability of the algorithm for quantum electron transport (in a far-from equilibrium time-dependent open system) is demonstrated by computing the current and fluctuations in a nano-resistor, with exchange and Coulomb interactions among electrons.
\end{abstract}

\maketitle

\section{Introduction}

A system with $N$ identical particles gives rise to a host of fascinating phenomena. Only those wave functions whose probability density remains unchanged under permutations of particles are a good description of such system. For separable Hamiltonians, these wave functions can be constructed from single-particle wave functions. However, for non-separable Hamiltonians, the computational burden associated with getting the $N$-particle wave function makes the exact solution inaccessible in most practical situations. This is known as the many-body problem \cite{Dirac}.

There has been a constant effort among the scientific community to provide solutions to the many-body problem. The quantum Monte Carlo solutions of the Schr\"{o}dinger equation provide approximate solutions to exact many-particle Hamiltonians \cite{nightingale99,suzuki93}.  The Hartree-Fock (HF) algorithm \cite{Hartree2,Fock} approximates the many-particle wave function by a single Slater determinant of non-interacting single-particle wave functions. Although it is known that the Hartee-Fock wave function cannot approach the original many-particle wave function, it can provide useful information on the original ground state. Alternatively, density functional theory (DFT) shows that the charge density can be used to compute any observable without the explicit knowledge of the many-particle wave function \cite{Nobel, hohenbergPR64}. Practical computations within DFT make use of the Kohn-Sham theorem \cite{kohn65PRA}, which defines a system of $N$ non-interacting single-particle wave functions that are able to provide a system of equations to find the exact charge density of the interacting system. However, the complexity of the many-body system is still present in the so called exchange-correlation functional, which is unknown and needs to be approximated. DFT has had a great success, mostly, in chemistry and material science \cite{solerJOPCM02}, both, dealing with equilibrium systems. Similar ideas can also be used for non-equilibrium time-dependent scenarios, through the Runge-Gross theorem \cite{rungePRL84}, leading to the time-dependent density functional theory (TDDFT). In contrast to the stationary-state DFT, where accurate exchange functionals exist, approximations to the time-dependent exchange-correlation functionals are still in their infancy.  TDDFT has been reformulated in terms of the current density \cite{giuliani05,TDDFTcurrent} and extended into a stochastic time-dependent current density when the system is interacting with a
bath \cite{ventraPRL07}.

The common strategy in all many-particle approximations is to obtain the observable result from mathematical entities defined in a real space, $\mathbf{R}^{3}$, (single-particle wave functions for HF and charge density for DFT) rather than from the many-particle wave function, whose support is defined in the configuration space $\mathbf{R}^{3N}$.

Bohmian mechanics \cite{HollandPR1993,llibrebohm,Bell1987,Durrllibre} is a consistent explanation of quantum phenomena based on the use of wave functions and trajectories. Apart from its ontological implications, Bohmian mechanics is nowadays used as a mathematical machinery that is able to reproduce the wave function evolution from fluid lines \cite{xoriols07PRL, waytt, bittner,tannor,meier}. This is the point of view used in this work to study exchange interaction in many-particle systems \cite{comment}. In Bohmian mechanics one can naturally find a single-particle wave function defined in $\mathbf{R}^{3}$, while still capturing many-particle features of the system. Such an entity is named conditional wave function \cite{llibrebohm}, and it is built by substituting all degrees of freedom present in the many-particle wave function, except one, by its corresponding Bohmian trajectories.  This substitution produces a single-particle wave function with a complicated time-dependence \cite{xoriols07PRL}. Recently, many-particle Bohmian trajectories associated to the conditional wave function have been investigated by Oriols et al. \cite{xoriols07PRL,galbareda08PRB} and the idea of introducing exchange interaction into non-separable systems through conditional wave functions was briefly indicated in the seminal work of  Ref. \cite{xoriols07PRL}.

The purpose of this paper is to present an algorithm to introduce exchange interaction into non-separable systems through conditional wave functions following the idea of Ref. \cite{xoriols07PRL}. This paper includes physical discussions, technical details and numerical results, omitted in Ref. \cite{xoriols07PRL},  that justifies the physical soundness of the proposal. The paper also includes the implementation of the exchange algorithm into a numerical simulator of quantum electron transport, justifying its numerical viability in practical systems. The paper is organized as follows.
In \sref{tra with wave} we provide an introduction to many-particle wave functions and Bohmian mechanics. For such introduction, we will use many-particle wave functions for separable Hamiltonians. From a didactic point of view,  these simple systems will be useful to discuss how the exchange interaction determines the behavior of the Bohmian trajectories of identical particles. In \sref{tra without wave}, we will explain how to compute many-particle Bohmian trajectories for identical particles for non-separable Hamiltonians, without computing the many-particle wave function.  Ensemble results for the kinetic, classical and quantum potential energies will be discussed for systems with and without exchange interaction. Finally, in \sref{alarcon_numerical_results}, we show the numerical viability of the algorithm to include exchange and Coulomb interaction for electron transport simulators. In \sref{Conclusions} we present the conclusions and some additional discussions.


\section{Many-particle trajectories from many-particle wave functions}
\label{tra with wave}

In this section, we introduce many particle wave functions and Bohmian mechanics to explain general properties of Bohmian trajectories associated to identical particles. These discussions will be of great utility in the subsequent sections.


\subsection{Summary of many-particle wave function}
\label{many wave}

For non-relativistic open systems of $N$-particles, a general expression for a many-particle wave function, $\Phi \equiv \Phi(\vec r_1,..,\vec r_N,t)$, with or without exchange interaction, is:
\begin{eqnarray}
\centering
\label{wave2}
\Phi = C \sum_{s_{zj}} \Psi&_{s_{z1},..,s_{zN}}(\vec{r}_{1},..,\vec r_N,t)
\gamma(s_{z1},..,s_{zN}),
\end{eqnarray}
where $\vec r_j$ represents the position of the $j$-th particle and $s_{zj}$ is the $z$-component of its spin, which can take the value $s_{zj}=\hbar/2$ (or $\uparrow_{j}$) for spin up and  $s_{zj}=-\hbar/2$ (or $\downarrow_{j})$ for spin down.  The normalization constant is $C$. The sum in  \eref{wave2} is over all possible combinations of spin \cite{notespin}.

In most discussions of this paper (except the numerical results discussed in \sref{alarcon_numerical_results}) we will assume that the quantum system is described by just one of the terms in \eref{wave2}. In particular, we will consider the term where all spins are parallel, e.g., $s_{zj}=\uparrow_{j}$ for $j=1,...N$.  In order to simplify our notation, the orbital part of this term will be written as $\Psi \equiv \Psi(\vec{r}_{1},..,\vec{r}_N,t)$, without any reference to the spins because their interchange becomes irrelevant.
Therefore, the (orbital) wave function  is solution of the following many-particle Schr\"{o}dinger equation:
\begin{eqnarray}
\centering
\label{manyscho}
i \hbar \frac{\partial \Psi}{\partial t} = \left( \sum_{k=1}^N -\frac{\hbar^2}{2m}  \nabla^2_k  + U(\vec r_{1},..,\vec r_{N},t) \right) \Psi,
\end{eqnarray}
where $m$ is the free electron mass and $U(\vec r_{1},\ldots,\vec r_{N},t)$ is a non-separable potential.
By construction, we know that the solution of \eref{manyscho} satisfies the following continuity equation:
\begin{equation}
\label{continuity}
\frac {d|\Psi|^2} {dt}+\sum_{k=1}^N \vec\nabla_{\vec r_k} \vec J_{\vec r_k} =0,
\end{equation}
where $\vec J_{\vec r_k} \equiv \vec J_{\vec r_k}(\vec r_1,...,\vec r_N,t)$ is the expectation values of the current probability density \cite{CohenTanudji} and $|\Psi|^2$ the presence probability density. This last result will be relevant in \sref{Many-particle trajectories} when presenting Bohmian trajectories.

Two particles are said to be identical if there are no experiments that can detect differences between them. This restriction on observable results can be satisfied by imposing the following property into the wave function  $\Psi$ of identical particles:
\begin{eqnarray}
\Psi(.,\vec r_j,.,\vec r_h,.,t) = e^{i \gamma}\Psi(.,\vec r_h,.,\vec r_j,.,t),
\label{exchange}
\end{eqnarray}
for any $j$ and $h$ indices. We consider $\gamma=0 \;(mod\;2\pi)$ for bosons (symmetry)  and $\gamma=\pi \;(mod\;2\pi)$ for fermions (antisymmetry).

We say that the system has exchange interaction when the wave function satisfies \eref{exchange}. For physical systems of identical particles, the many-particle potential in \eref{manyscho} remains invariant under the permutation of two positions, i.e. $U(.,\vec r_{j},...,\vec r_{h},.,t)=U(.,\vec r_{h},...,\vec r_{j},.,t)$ for any $j$ and $h$, and the symmetry or antisymmetry property of the wave function in \eref{exchange} for time $t$ holds for all instants. Next, as a simple example of the difference between systems with and without exchange interaction, we discuss on the total energy, which will be useful later in \sref{alarcon_numerical_results}.


\subsubsection{Example: The effect of exchange interaction on total energy}
\label{total energy}

We consider a system of  $N$ particles in free space. For simplicity, we consider 1D particles where its position is defined in $\mathbf{R}$. Then, the many-particle wave function $\Psi(x_1,...,x_N,0)$ at $t=0$ can be constructed from the following single-particle Gaussian wave packets:
\begin{eqnarray}
\psi _j (x_j,0 ) = \frac{\exp { \left( {i k_{oj} x_j } \right)}}{{\left( {\pi \sigma _{x_j }^2 } \right)^{1/4} }} \exp{\left( { - \frac{{\left( {x_j  - x_{oj} } \right)^2 }}{{2\sigma _{x_j }^2 }}} \right)},
\label{gausiana}
\end{eqnarray}
where $\sigma _{xj} $ is the spatial dispersion, $x_{oj}$ the central position, $E_{oj}=(\hbar \; k_{oj})^2/(2 \;m)$ the central energy of each wave packets and $k_{oj}$ the central wave vector.

In particular, the $N$-particle wave function $\Psi(x_1,...,x_N,0)$ with exchange interaction can be defined from:
\begin{eqnarray}
\Psi(x_{1},.., x_{N},0) = &C& \; \sum_{n=1}^{N!}\prod_{j=1}^{N}  \psi _j(x_{p(n)_j},0) \; {\rm sign}(\vec{p}_{n}),
\label{slater}
\end{eqnarray}
where the sum is over all $N!$ permutations $\vec{p}_{n}=\{p(n)_{1},...,p(n)_{N}\}$ and  $C$ is a normalization constant.  For fermions, the ${\rm sign}(\vec{p}_{n})= \pm 1$ means the sign of the permutations, i.e. \eref{slater} is the Slater determinant. Alternatively, we will consider ${\rm sign}(\vec{p}_{n})= 1$ for bosons, meaning that \eref{slater} has to be interpreted as the permanent.

On the other hand, the wave function for particles without exchange interaction can be written as:
\begin{equation}
\Psi(x_{1}, ..., x_{N},0)= \prod_{j=1}^{N}  \psi _j(x_j,0),
\label{noslater}
\end{equation}
which, by construction, is already well normalized to unity. The ensemble value of the kinetic energy of the $j$-th particle belonging to a system of particles without exchange interaction is computed as:
\begin{equation}
\left\langle T_j \right\rangle  = \int ... \int {\Psi^{*}{\rm{ }} \hat{T}_j \Psi \; dx_1 .. dx_N } ,
\label{kinetic energy}
\end{equation}
The kinetic energy operator is $\hat{T}_j ={ - \frac{{\hbar ^2 }}{{2m}}\frac{{\partial ^2 }}{{\partial x_j^2 }}}$. Hereafter, unless specified, the spatial integrals are assumed to extend over the whole configuration space. The same expression \eref{kinetic energy} can be used for identical particles defined from the wave function in \eref{slater}. Then, one can easily realize that $\left\langle T_{j} \right\rangle=\left\langle T_{h} \right\rangle$ for any $j$ and $h$ indexes. As expected, one cannot discern between identical particles from the measurement of their kinetic energies.

\begin{figure}[h]
\begin{center}
\includegraphics*[width=9cm,height=7cm]{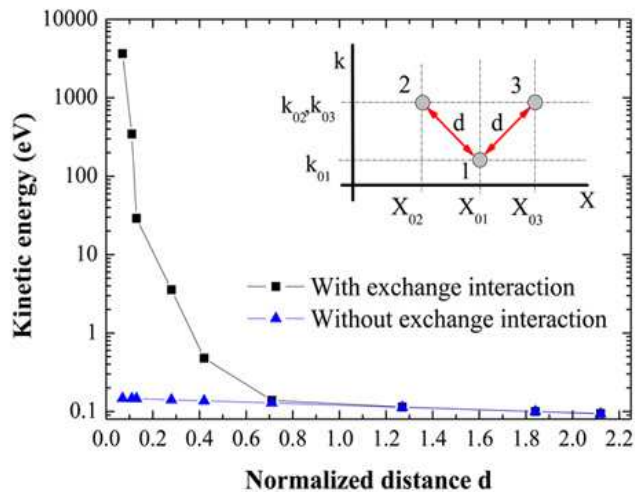}
\caption{%
  \footnotesize{ (Color online) Ensemble value of the kinetic energy for a 3-particle system with (square solid black line) and without (up triangle solid blue line) exchange interaction as function of their normalized phase-space distance $d$. The inset shows the positions of the central position and central wave vector of each wave packet in the phase-space, which are used to define the distance $d$ among them. }}
\label{Fig_ec_d}
\end{center}
\end{figure}

We compute the behavior of the total kinetic energy  $\left\langle T \right\rangle=\left\langle T_{1} \right\rangle+\left\langle T_{2} \right\rangle+\left\langle T_{3} \right\rangle$ for three electrons (with parallel spins) with and without exchange interaction, as a function of the distance among the wave packets in the configuration space (see inset in \fref{Fig_ec_d}). We define the normalized phase-space distance among the central positions and central wave vectors of two wave packets as \cite{xoriols04Nano}:
\begin{equation}
d(1,j)^2  = \frac{{(k_{o1}  - k_{oj} )^2 }}{{2\sigma _k^2 }} + \frac{{(x_{o1}  - x_{oj} )^2 }}{{2\sigma _x^2 }};\;\;j=2,3 ,
\label{distance}
\end{equation}
where $\sigma _{kj}  = {1}/{{\sigma _{xj} }}$ is the wave vector dispersion. In \fref{Fig_ec_d}, we plot,  in a square (black) line,  the mean value of the total kinetic energy of three electron (fermions) with exchange interaction, whose wave function is defined from \eref{slater}. The result is repeated  for different values of the  distance $d=d(1,2)=d(1,3)$ with the condition $x_{o1}-x_{o2}=x_{o3}-x_{o1}$ and $k_{o2}-k_{o1}=k_{o3}-k_{o1}$ seen in the inset of \fref{Fig_ec_d}. Identically, we plot in up triangle (blue) line the total kinetic energy computed for three particles without exchange interaction, whose wave function is defined from \eref{noslater}. For large $d$, the values of the kinetic energy of the three electrons with and without exchange interaction are identical. For such large values of $d$,  all electrons are placed far away from each other in the phase-space and the exchange interaction has no effect. However, this is not true for small values of $d$. Then,  the difference between the kinetic energy of electrons with or without exchange interaction increases as we place the three electrons closer inside the phase-space.

The result plotted in \fref{Fig_ec_d} is just (the wave packet version of) the celebrated Pauli exclusion principle: identical fermions cannot be in the same quantum state. The discussion has been done with three particles, instead of two, because in \ref{appendix2} we generalize the present example to three electrons with exchange interaction and different spins orientations.


\subsection{Summary of many-particle trajectories}
\label{Many-particle trajectories}


In Bohmian mechanics \cite{HollandPR1993,llibrebohm,Bell1987,Durrllibre},  each particle of the system is represented by a trajectory guided by a wave. The wave is the many-particle wave function discussed above, $\Psi (\vec r_1,...,\vec r_{N},t)$, with all its computational difficulties. Such wave function satisfies the continuity equation, written in \eref{continuity}, that relates current and probability presence densities. From such continuity equation, one can easily define a (Bohmian) velocity  $\vec v_j(\vec r_1,...,\vec r_N,t)$ at each position of the configuration space as:
\begin{equation}
\vec v_j(\vec r_1,...,\vec r_N,t)=\frac {\vec J_{\vec r_j}(\vec r_1,...,\vec r_N,t)} {|\Psi(\vec r_1,...,\vec r_N,t)|^2}.
\label{Bohmianvelocity}
\end{equation}
The (Bohmian) trajectory of the $j$-th particle, $\vec r_j^l[t]$, in real space can be defined by time-integrating \eref{Bohmianvelocity} as:
\begin{equation}
\vec r_j^l[t]=\vec r_j^l[0]+\int_{0}^{t} \vec v^l_j(\vec r_1^{l}[t\rq{}],...,\vec r_N^{l}[t\rq{}],t\rq{}) dt\rq{}.
\label{trajectory}
\end{equation}
Obviously, one has to select the initial position $\vec r_j^{l}[0]$ to perfectly specify the trajectory. The super-index $l={1,...,M}$ on the trajectory accounts for the $M\rightarrow\infty$ different initial positions that can be selected. We refer to $\vec r_j^l[t]$ as the Bohmian trajectory in $\mathbf{R}^{3}$, while we will refer to $\{\vec r_1^l[t],..,\vec r_N^l[t]\}$ as a many-particle (or $N$-particle) Bohmian trajectory in $\mathbf{R}^{3N}$.

The relevant property of these Bohmian trajectories that makes them meaningful for quantum computations is the fact that, by construction, a proper ensemble of them (with different initial positions) does exactly reproduce the time-evolution of the many-particle wave function, at any time. A proper ensemble means that the initial positions, $\{\vec r_1^{l}[0]....\vec r^{l}_{N}[0] \}$, are selected according to the probability distribution $|\Psi (\vec r_1,...,\vec r_{N},0)|^2$. This last condition is called 'quantum equilibrium hypothesis' \cite{HollandPR1993,goldstein}.

We can now deduce an important property of these trajectories that will be very relevant later. Since $\Psi (\vec r_1,...,\vec r_{N},t)$ is a single-valued wave function, the Bohmian velocity computed from  \eref{Bohmianvelocity} in each point of the configuration space is unique. This means that if two trajectories coincide at some point of the configuration space, then, they will coincide forever (because their velocities become identical). This well-known result can be summarized in a simple sentence: two many-particle Bohmian trajectories (with different initial positions) do not cross in the configuration space, either for bosons, fermions or non-identical particles \cite{pracros}.

Equivalently, the presentation of such trajectories can be done by introducing the polar form of the many-particle wave function $\psi(\vec r_1,..,\vec r_{N},t) = R(\vec r_1,..,\vec r_{N},t) e^{i S(\vec r_1,..,\vec r_{N},t)/\hbar}$ into \eref{manyscho}. The modulus $R \equiv R(\vec r_1,..,\vec r_{N},t)$ and the phase $S \equiv S(\vec r_1,..,\vec r_{N},t)$ are real functions. Then, one obtains again, from the imaginary part of \eref{manyscho}, the continuity equation defined in (\ref{continuity}) in polar form:
\begin{eqnarray}
\label{charge_conservationND}
\frac{\partial R^{2}}{\partial t}+ \sum_{j = 1}^{N} \vec \nabla_{\vec r_j} \left( R^2 \frac {\vec \nabla_{\vec r_j} S} {m}  \right) = 0,\nonumber \\
\end{eqnarray}
where we recognize the velocity of the $j$-th particle as:
\begin{equation}
\label{Bohmianvelocity2}
\vec v_j(\vec r_1,..,\vec r_{N},t) = \frac {\vec \nabla_{\vec r_j} S(\vec r_1,..,\vec r_{N},t)} {m} .
\end{equation}
By construction \cite{llibrebohm}, the velocity definition in \eref{Bohmianvelocity2} is identical to that in \eref{Bohmianvelocity}. On the other hand, the real part of the Schr\"odinger equation leads to a many-particle version of the quantum Hamilton--Jacobi equation:
\begin{eqnarray}
\label{Hamilton_JacobiND}
\frac{\partial S}{\partial t} + U + \sum_{j = 1}^{N} \left(K_j  + Q_j \right)
  = 0,
\end{eqnarray}
where $U \equiv U(\vec r_1,..,\vec r_{N},t)$ is the potential in \eref{manyscho} and we have defined the (local) Bohmian kinetic energy as:
\begin{equation}
\label{bohmiankinetic}
K_j \equiv K_j(\vec r_1,..,\vec r_{N},t) = \frac {1} {2} m \vec v_j(\vec r_1,..,\vec r_{N},t)^2,
\end{equation}
and the (local) quantum potential energy:
\begin{equation}
\label{quantumpotential}
Q_j \equiv Q_j(\vec r_1,..,\vec r_{N},t) = -\frac{\hbar^2} {2 m} \frac{\vec \nabla_{\vec r_j}^2 R(\vec r_1,..,\vec r_{N},t)} {R(\vec r_1,..,\vec r_{N},t)}.
\end{equation}
When dealing with Bohmian trajectories, the ensemble kinetic energy defined in \eref{kinetic energy} is divided into two parts, $\left\langle \hat{T}_j \right\rangle=\left\langle \hat{K}_j \right\rangle+\left\langle \hat{Q}_j \right\rangle$. The first part:
\begin{eqnarray}
\left\langle \hat{K}_j \right\rangle = \int ... \int R^2 \; K_j \; dx_1...dx_N,
\label{enkinetic}
\end{eqnarray}
related to the local (Bohmian) kinetic energy $K_j \equiv K_j(x_1,...x_N,t)$, and the second part:
\begin{eqnarray}
\left\langle \hat{Q}_j \right\rangle = \int ... \int R^2 \; Q_j \; dx_1...dx_N,
\label{enquantum}
\end{eqnarray}
to the quantum potential energy $Q_j \equiv Q_j(x_1,...,x_N,t)$.


\subsubsection{Properties of many-particle Bohmian trajectories with exchange interaction}
\label{properties_trajectories}

Now, we can list a series of important properties for those ensembles of Bohmian trajectories that represents identical particles, i.e., when exchange interaction is present. In order to simplify the notation, we define $\vec X=\{\vec r_1,..,\vec r_{N}\}$. Identically, we define the $N$-particle Bohmian trajectory at time $t=0$ as $\vec X^l[0]=\{\vec r_1^l[0],..,\vec r_{N}^l[0]\}$. Another set of initial conditions will be refereed as $\vec X^f[0]=\{.,\vec r_h^f[0],.,\vec r_{j}^f[0],.\}$ when it contains the same initials positions as $\vec X^l[0]$, but the two initial positions, $\vec r_j^l[0]$ and $\vec r_h^l[0]$, are interchanged. Because of \eref{exchange}, the modulus of the many-particle wave function satisfies:
\begin{eqnarray}
R(\vec X^l[0],0) = R(\vec X^f[0],0),
\label{exchangeR}
\end{eqnarray}
for any such type of two set of initials conditions $l$ and $f$. Identically, the phase satisfies:
\begin{eqnarray}
S(\vec X^l[0],0) =\gamma+ S(\vec X^f[0],0),
\label{exchangeS}
\end{eqnarray}
where $\gamma=0 \;(mod\;2\pi)$ for bosons (symmetry)  and $\gamma=\pi \;(mod\;2\pi)$ for fermions (antisymmetry). As discussed for the wave function, the requirements in \eref{exchangeR} and \eref{exchangeS} are satisfied at any time $t$. The property of \eref{exchangeS} togther with the definition of the velocity in \eref{Bohmianvelocity2} implies:
\begin{eqnarray}
\vec v_j(\vec X^l[t],t) = \vec v_h(\vec X^f[t],t).
\label{exchangevelo}
\end{eqnarray}
This condition on the Bohmian velocities, which is valid for either bosons or fermions, has two relevant consequences. First, let us compare the two sets of many-particle trajectory with different initial positions mentioned above: the $l$-set and the $f$-set. Their difference are only $\vec r_j^l[0]=\vec r_h^f[0]$ and $\vec r_h^l[0]=\vec r_j^f[0]$. Then, we realize from \eref{exchangevelo} that all Bohmian trajectories with identical initial conditions will be equal independently of the initial conditions, except the two trajectories which have their initial positions interchanged. For these trajectories, we get $\vec r_j^l[t]=\vec r_h^f[t]$ and $\vec r_h^l[t]=\vec r_j^f[t]$.

The second consequence of \eref{exchangevelo} is valid for those many-particle trajectories that have, at least, two equal components, i.e.  $\vec r_j^l[0]=\vec r_h^l[0] \equiv \vec a$. Because of this coincidence, we have $\vec X^l[0]=\{.,\vec a,.,\vec a,.\}$ and also $\vec X^f[0]=\{.,\vec a,.,\vec a,.\}$ which in fact are the same. Then, the condition $\vec v_j(\vec X^l[t],t) = \vec v_h(\vec X^f[t],t)$ can be written as:
\begin{eqnarray}
\vec v_j(\vec X^l[t],0)= \vec v_h(\vec X^l[t],0) \equiv v_{a},
\label{velodiagonal}
\end{eqnarray}
because $\vec X^l[0]=\vec X^f[0]$. Then, the trajectory $\vec r_j^l[t]$ at the subsequent time $\vec r^l_j[0+dt]=\vec a+\vec v_a dt$ is identical to the other trajectory $\vec r_h^l[0+dt]=\vec a+\vec v_a dt$. This result, means $\vec r_j^l[t]=\vec r_h^l[t]$ at any time.

 Because of the previous property and the non-crossing property of Bohmian trajectories discussed before \cite{pracros}, we have an important corollary. We define \lq\lq{}diagonal\rq\rq{} many-particle trajectories as those trajectories where at least two components, $\vec r_j^l[t]=\vec r_h^l[t]$, are identical (the rest of components can be different). Since other Bohmian trajectories cannot cross such \lq\lq{}diagonal\rq\rq{} trajectories, all Bohmian trajectories are restricted  to remain in subspaces of the configuration space. According to Ref. \cite{Baci},  Bohmian mechanics for identical particles can be described in a  "reduced" space $\mathbf{R}^{3N}/S_N$, with $S_N$ the permutation space of $N-$ particles.

Finally, we want to mention that  in Bohmian computations, even with the symmetrization postulate, trajectories of particles are obviously distinguishable. One labels the trajectory of particle $1$ as $\vec r_1^l[t]$ and that of particle $2$ as $\vec r_2^l[t]$. We have shown that, by construction,  the Bohmian trajectories have special symmetry requirements. Then, all results for particle $1$ computed from an ensemble of these trajectories will be identical to those computed for particle $2$. In simple words, for a system of identical particles, Bohmian trajectories are
distinguishable, while observable results associated to different particles become indistinguishable.


\subsubsection{Example: The effect of exchange interaction on Bohmian trajectories}
\label{Bohmian_trajectories_example}


Let us discuss, with some numerical examples, the previous properties of Bohmian trajectories of identical particles.
In all the numerical examples of this subsection, we consider two free particles propagating, each one, in 1D physical space. The single-particle wave packets that will be used to construct the many-particle wave function at the initial time $t=0$ are defined from \eref{gausiana}.

\begin{figure}
\centering
\includegraphics*[width=8cm]{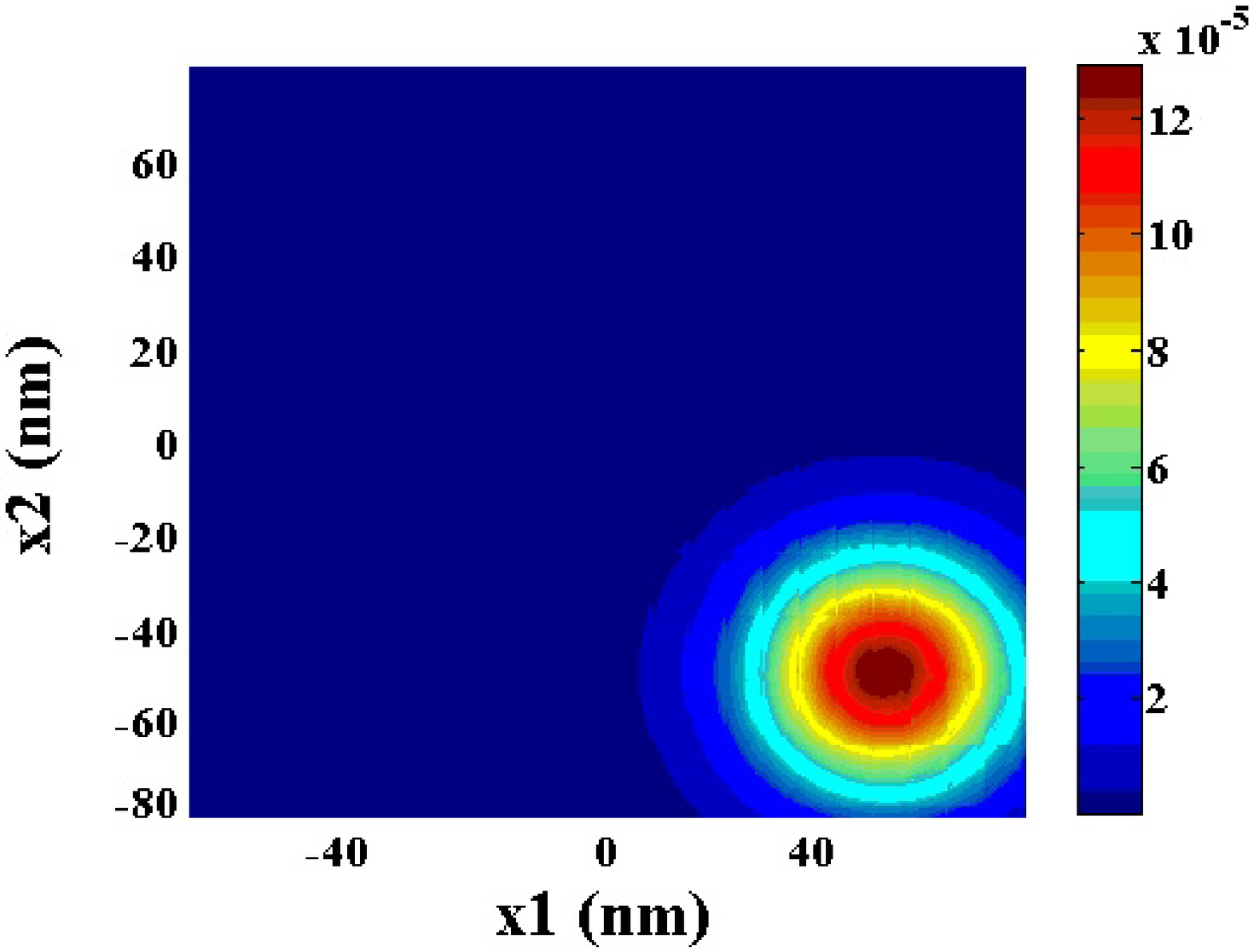}
\caption{%
  \footnotesize{ (Color online) Modulus of the wave function for two particles without exchange interaction in the 2D configuration space at $t=0$ fs.}}
\label{Fig2}
\includegraphics*[width=8cm]{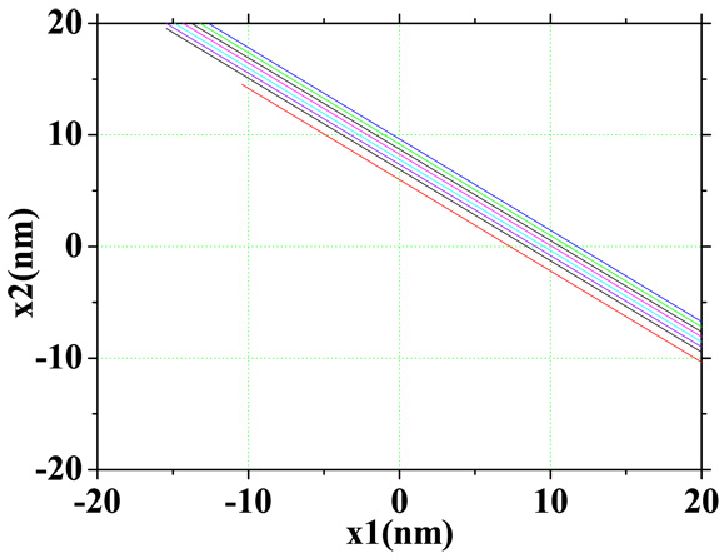}
\caption{%
  \footnotesize{(Color online) Two-particles Bohmian trajectories with different initials conditions for particles without exchange interaction in a free space.}}
\label{Fig2b}
\includegraphics*[width=8cm]{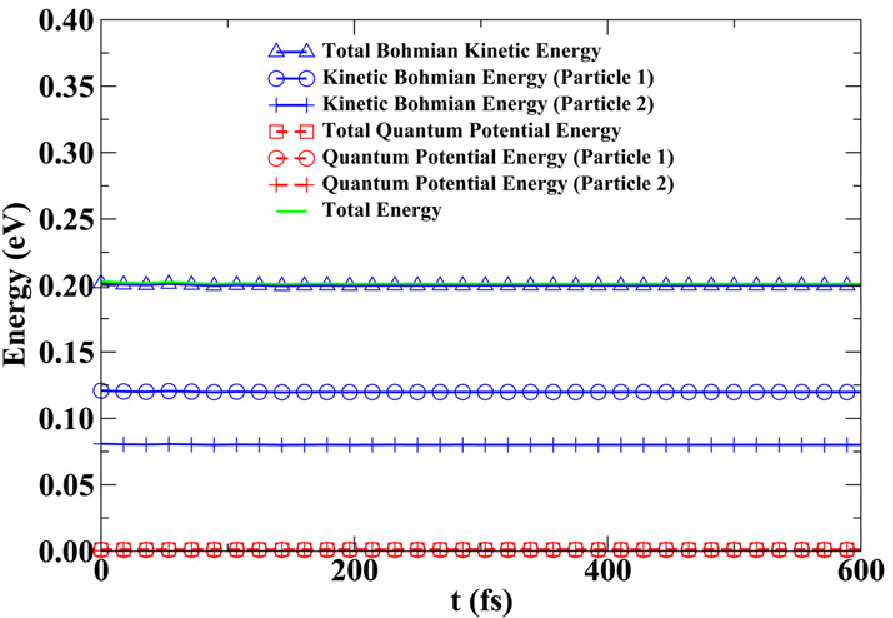}
\caption{%
  \footnotesize{ (Color online) Time evolution of the total and individual (ensemble average) energies of two-electron system without exchange interaction in free space. }}
\label{Fig3}
\end{figure}

First, we consider two electrons with a wave function $\Psi (x_1,x_2,t)$ computed from \eref{noslater}, without any symmetry. See the initial modulus of the 2-particle wave function in \fref{Fig2}. In particular, we consider $E_{o1}=0.12$ eV, $x_{o1}=+50$ nm and $\sigma_{x1}=25$ nm for the first wave packet, and $E_{o2}=0.08$ eV, $x_{o2}=-50$ nm and $\sigma_{x2}=25$ nm for the second. In order to see the spatial interaction of the two particles, the momentum of the first particle is negative and that of the second positive.  We consider a free electron mass for both electrons. Once we know $\Psi (x_1,x_2,t)$,  we compute the (two-particle) Bohmian trajectory from \eref{trajectory} with different initial positions. As seen in \fref{Fig2b},
they correspond to roughly parallel lines.

In \fref{Fig3}, we see that the quantum potential of either the first or the second particles are nearly zero. The total quantum potential, as the sum of the two particles, is plotted in dashed square (red) line. The ensemble (Bohmian) kinetic energy remains equal to its initial value, $0.12$ eV for the first wave packet in solid circle (blue) line and $0.08$ eV for the second in solid plus (blue) line. The total energy in solid (green) line remains constant and equal to $0.2$ eV, i.e., the sum of kinetic energies. These simple Bohmian trajectories for non-identical particles move roughly like classical particles.

\begin{figure}
\begin{minipage}{15.8cm}
\centering
\includegraphics*[width=7.5cm]{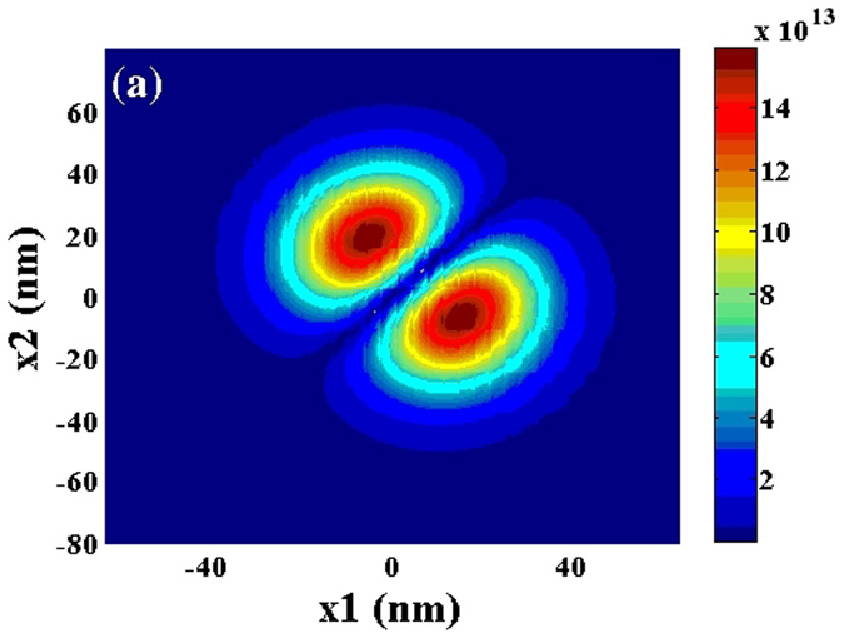}
\includegraphics*[width=7.5cm]{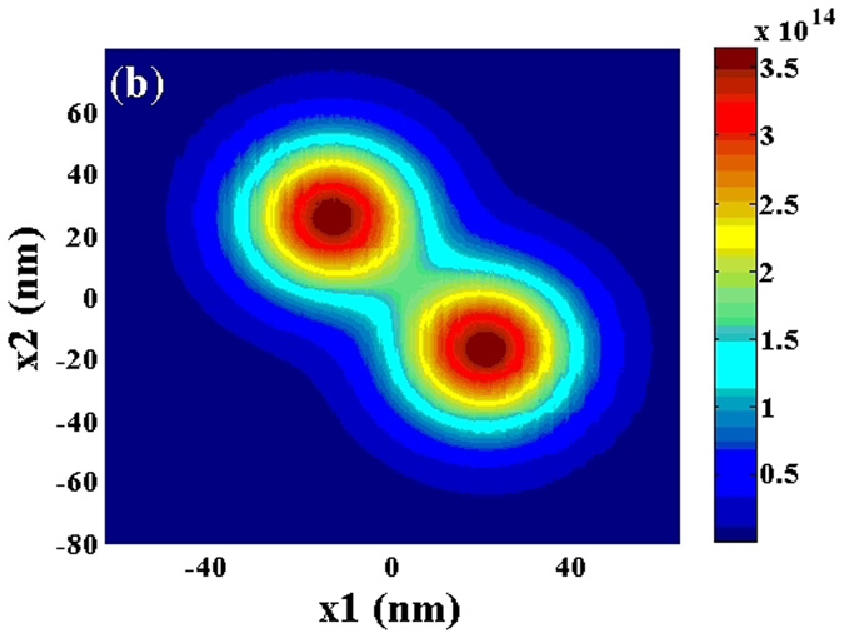}
\caption{%
  \footnotesize{ (Color online) Modulus of the wave function for two identical particles in the 2D configuration space. (a) two fermions at $t=267.8$ fs and (b) two bosons at $t=178.5$ fs.}}
\label{Fig5}

\includegraphics*[width=7.5cm]{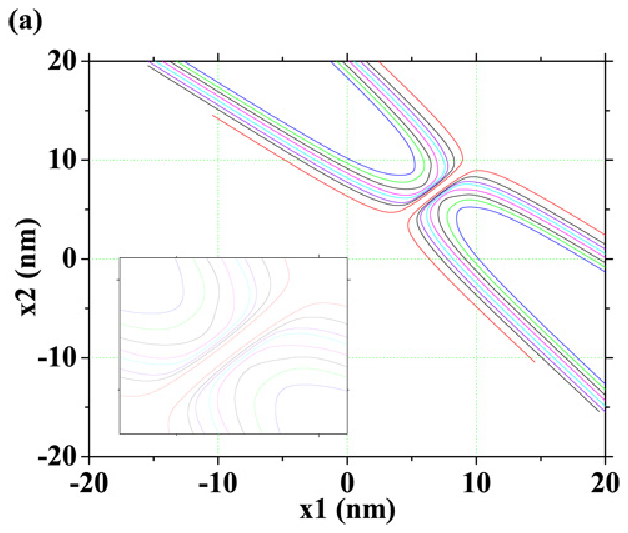}
\includegraphics*[width=7.5cm]{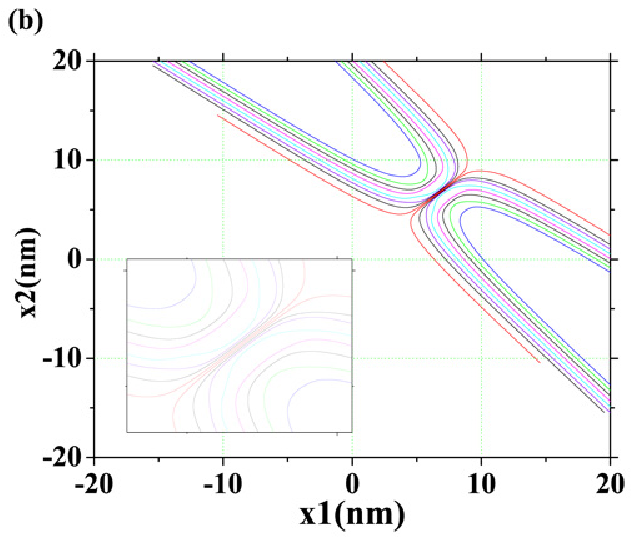}
\caption{%
  \footnotesize{ (Color online) Two-particle Bohmian trajectories with different initial conditions in free space. (a) Fermions and (b) bosons. The inset is a zoom of the diagonal non-crossing properties of Bohmian trajectories.}}
\label{Fig6}

\includegraphics*[width=7.5cm]{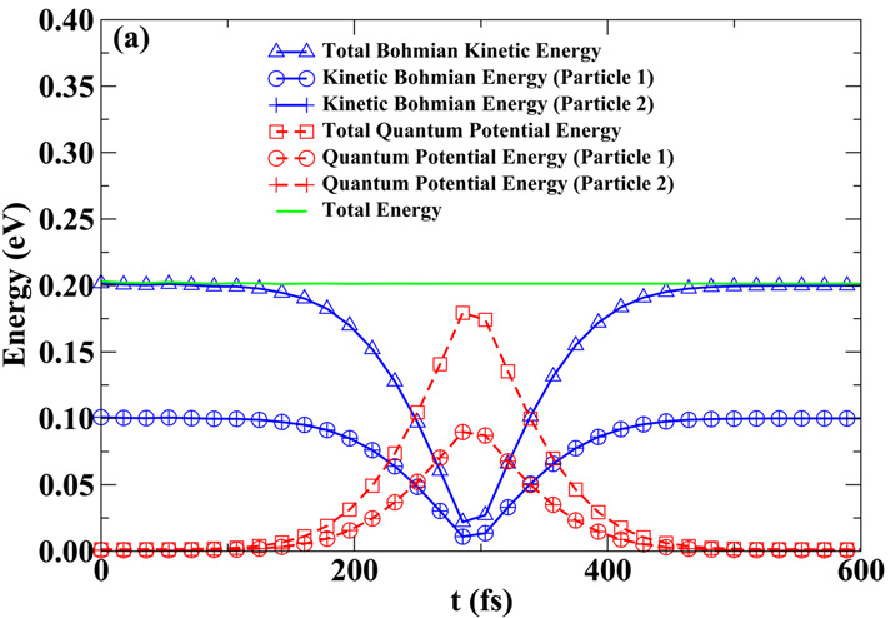}
\includegraphics*[width=7.5cm]{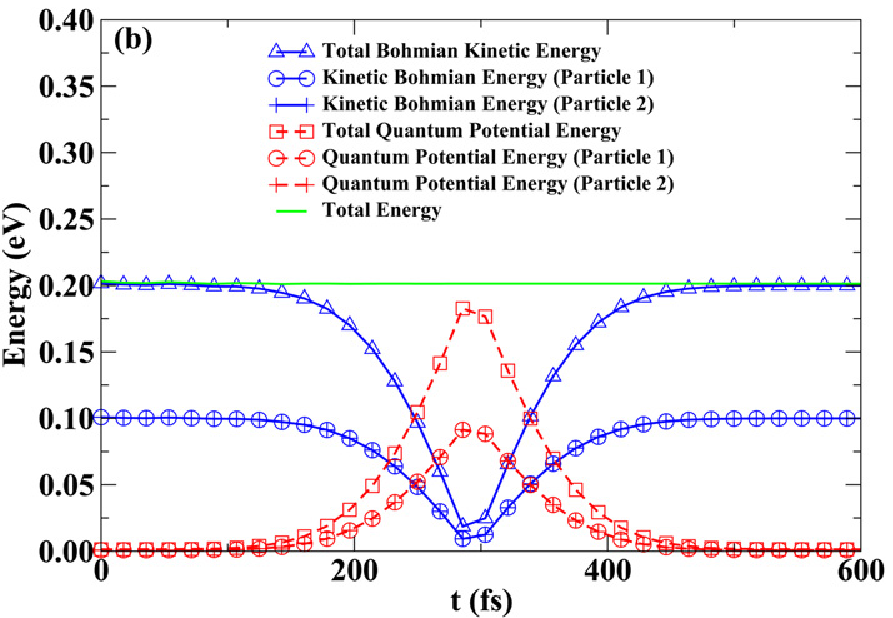}
\caption{%
  \footnotesize{(Color online) Time evolution of the total and individual (ensemble average) energies of the two-particle system in a free space. (a) Fermions and (b) bosons.\vspace{2cm} }}
\label{Fig7}

\end{minipage}
\end{figure}

Next, we consider a wave function $\Psi (x_1,x_2,t)$ of two identical electrons computed from the Slater determinant of \eref{slater} for $N=2$. We use the same two initial gaussian wave packets discussed above. In \fref{Fig5}(a), we plot the (symmetric) modulus of the many-particle wave functions at $t=267.8$ fs. In particular, we get $\Psi (a,a,t)=0$ at any point $\{a,a\}$ of the diagonal. In \fref{Fig6}(a), we plot a set of Bohmian trajectories. The initials positions $\{x_1^l[0],x_2^l[0]\}$ are selected symmetrically with respect to the \lq\lq{}diagonal\rq\rq{}. First, we observe that $x_1^l[t]=x_2^f[t]$  and $x_2^l[t]=x_1^f[t]$ when $x_1^l[0]=x_2^f[0]$ and $x_2^l[0]=x_1^f[0]$. As discussed in \sref{properties_trajectories}, the Bohmian trajectories corresponding to interchanged initial positions become symmetrical with respect to the diagonal points of the configuration space.  Second, we observe that the Bohmian trajectories do not cross the diagonal.

In \fref{Fig7}(a), we plot the energies of this two-particle fermion system. The total energy of the identical particles is equal to that of the particles without exchange interaction discussed in \fref{Fig3}. The reason, as explained in \sref{total energy}, is because the momentum of the wave packets are very different. One momentum is positive and the other negative and no Pauli effect is observed in the energy. However, since Bohmian trajectories are ``reflected'' at the diagonal, their (bohmian) velocity becomes zero at that time. Then, the ensemble average of $K_j(x_1,x_2,t)$ in \eref{enkinetic} is almost zero, while the ensemble average of $Q_j(x_1,x_2,t)$ grows to keep the total energy constant. The same result can be argued by noting that the quantum potential in \eref{quantumpotential} depends on the curvature of the modulus, which becomes large at that points. In addition, in contrast to the two particles without exchange interaction discussed in \fref{Fig3}, the (Bohmian kinetic plus quantum) energies of the first particle are identical to those of the second particle. The observable results of the energy of the individual particles are indistinguishable, while we can perfectly distinguish the trajectories in \fref{Fig5}(a).

Finally, in \fref{Fig5}(b), \fref{Fig6}(b) and \fref{Fig7}(b),  we plot the same result as in the previous figures but considering two identical bosons. We use exactly the same parameters for the wave packets discussed in the previous figures. The only difference is that the initial wave function is computed from \eref{slater} when the ${\rm sign}(\vec {p}_{n})$ is substituted by $1$. Let us notice again the symmetric property of the modulus of the wave function in the configuration space. Although we have $\Psi (a,a,t) \neq 0$ at the diagonal points $\{a,a\}$, we see in \fref{Fig6}(b), that Bohmian trajectories do not cross that diagonal. This is an expected result because our discussions on the properties of Bohmian trajectories in \sref{properties_trajectories} do not depend on the bosonic or fermionic nature of particles. There is a Bohmian trajectory located along the diagonal points of the configuration space (not plotted) that does not allow to be crossed by other trajectories. The initials positions $\{x_1^l[0],x_2^l[0]\}$ are selected symmetrically with respect to the \lq\lq{}diagonal\rq\rq{} and identical to the ones used for the fermions. Again, trajectories are symmetric under the exchange of initial positions. In \fref{Fig7}(b), we plot the energies of the two-particle bosonic system. The numerical results of the (Bohmian) kinetic energy for bosons (0.019 eV) are slightly lower than fermions (0.022 eV) when the wave packet is close to the diagonal of the configuration space. The reason is because there are more bosonic Bohmian trajectories that arrive closer to the diagonal in \fref{Fig5}(b) and \fref{Fig6}(b) than the fermionic ones in \fref{Fig5}(a) and \fref{Fig6}(a).

\section{Many-particle trajectories without many-particle wave functions}
\label{tra without wave}


As commented in the introduction, many attempts have been developed in the literature to provide accurate solutions to the many-body problem. Here, we briefly review one of this approximations presented by one of the authors in  Ref. \cite{xoriols07PRL}. Then, we explain how the exchange interaction can be included in the mentioned approximation.


\subsection{The conditional wave function}
\label{review}


The main idea behind the many-body approximation mentioned in Ref. \cite{xoriols07PRL} is the fact that the computation of the Bohmian velocity for the $\vec r_a[t]$ trajectory from \eref{Bohmianvelocity} only requires the spatial derivatives of $\Psi(\vec r_1,..,\vec r_a,..,\vec r_N,t)$ on the $\vec r_a$ directions, and not on the rest of degrees of freedom. Thus, in principle, the trajectory $\vec{r}_a[t]$ can be equivalently computed from  the many-body wave function $\Psi(\vec r_1,...,\vec r_N,t)$ or from the following conditional wave function:
\begin{eqnarray}
\label{wavecon}
&\Psi&_{a}(\vec{r}_a, t) = \Psi(\vec r_a,\vec X_{a}[t],t),
\end{eqnarray}
where $\vec{X}_{a}[t]=\{\vec{r}_{1}[t],\vec{r}_{a-1}[t],\vec{r}_{a+1}[t],\vec{r}_{N}[t]\}$ is a vector that contains all Bohmian trajectories except  $\vec{r}_a[t]$. We also use $\vec X_{a}=\{\vec r_1,...\vec r_{a-1},\vec r_{a+1},..,\vec r_N\}$ when referring to all the degrees of freedom except $\vec r_a$. When not relevant, we avoid the superindex $l$ in the Bohmian trajectory that specifies the initial positions of the trajectory. Certainly, the conditional wave function  in \eref{wavecon} is defined in a much smaller configuration space, $\mathbf{R}^{3}$,  than the many-body wave function. Thus, in principle, the conditional wave function needs much less computational effort than the explicit  many-particle wave function.
Following Ref. \cite{xoriols07PRL}, the single-particle wave function $\Psi_{a}(\vec{r}_a, t)$, that we will use to compute $\vec r_a[t]$, can be obtained as a solution of the single-particle Schr\"{o}dinger equation:
\begin{eqnarray}
i \hbar\frac{\partial \Psi_{a}(\vec{r}_{a},t)}{\partial t} = \left( -\frac{\hbar^2}{2m}\nabla^{2}_{\vec{r}_{a}}
+ U_{a}(\vec{r}_{a},\vec{X}_{a}[t],t)+G_{a}(\vec{r}_{a},\vec{X}_{a}[t],t)\right. \; \nonumber \\
\left.+ i J_{a}(\vec{r}_{a},\vec{X}_{a}[t],t) \right) \Psi_{a}(\vec{r}_{a},t).
\label{pseudo}
\end{eqnarray}
The exact definition of the terms $U_{a}(\vec{r}_{a},\vec{X}_{a},t)$, $G_{a}(\vec{r}_{a},\vec{X}_{a},t)$ and $J_{a}(\vec{r}_{a},\vec{X}_{a},t)$ can be found in Ref. \cite{xoriols07PRL}.

In brief,  we have been able to decompose an irresolvable $N$-particle Schr\"{o}dinger equation into a set of $N$-single-particle Schr\"{o}dinger equation with time-dependent potentials \cite{xoriols07PRL}. At this point we realize that the extraordinary numerical simplification comes at the prize that there are terms in \eref{pseudo} which are unknown and need pertinent approximations, $G_{a}(\vec{r}_{a},\vec{X}_{a}[t],t)$ and $J_{a}(\vec{r}_{a},\vec{X}_{a}[t],t)$. This is a similar situation to that in DFT discussed in the introduction.

\subsubsection{Test for non-separable harmonic potentials without exchange interaction}
\label{Moises_numerical_nonidentical}


Next, in order to clarify the use of conditional (Bohmian) trajectories discussed above, we applied it to a simple system of two electrons without exchange interaction under a non-separable Hamiltonian. We consider two 1D particles so that the configuration space is $\mathbf{R}^{2}$. We use the non-separable potential energy:
\begin{eqnarray}
\label{coulomb2}
U(x_{1},x_{2})=c (x_1-x_2)^2,
\end{eqnarray}
where the factor $c$ will allow us to modify arbitrarily the strength of the non-separable interaction. In particular, we will use $c=10^{12} \;eV/m^2$.  The  many-body wave function $\Psi(x_{1},x_{2},t)$ can be solved exactly from \eref{manyscho} with $N=2$. Once the exact 2D wave function $\Psi(x_{1},x_{2},t)$ is known, we can compute the exact 2D Bohmian trajectories straightforwardly from \eref{Bohmianvelocity}.

\begin{figure}[ht]%
\centering
\includegraphics[width=8cm]{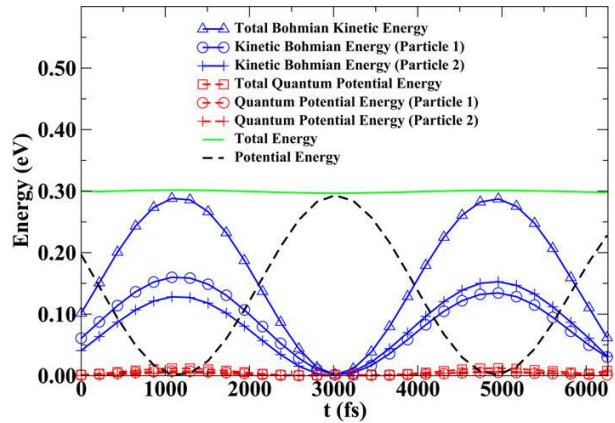}
\caption{ \footnotesize{(Color online) Time evolution of the total and individual (ensemble averaged) energies of two-electron system without exchange interaction under a non-separable potential.  }}
\label{fig11}
\end{figure}

In \fref{fig11}, we have plotted the ensemble results of the (Bohmian) kinetic energy, \eref{enkinetic}, the quantum  potential energy, \eref{enquantum}, for the two electrons.  We compute the results directly from the 2D exact wave function solution of \eref{manyscho}. We emphasize that there is an interchange of kinetic energies between the first and second particles (see their kinetic energy in the first and second oscillations). This effect clearly manifests that the Hamiltonian of that quantum system is non-separable.

Alternatively, we can compute the trajectories used to compute \fref{fig11} without knowing the many-particle wave function, but computing the conditional wave function $\Psi_a(x_a,t)$ solution of \eref{pseudo} with the proper approximation for terms $G_{a}$ and $J_{a}$. Here, we consider a zero order Taylor expansion around $x_a[t]$ for the unknown potentials terms $G_{a}$ and $J_{a}$.  In other words, we consider them as purely time-dependent potential terms, $G_{a}(x_{a},x_{b}[t],t) \approx G_{a}^{''}(x_a[t],t)$ and $J_{a}(x_{a},x_{b}[t],t) \approx J_{a}^{''}(x_a[t],t)$. This is the simplest approximation. Then, we know that the (complex) purely time-dependent terms $G_{a}^{''}(x_{a}[t],\vec{X}_{a}[t],t)$ and $J_{a}^{''}(x_{a}[t],\vec{X}_{a}[t],t)$ in the Hamiltonian of \eref{pseudo} only introduce a (complex) purely time-dependent phase. Then, we can write $\Psi_a(x_a,t)$ as:
\begin{eqnarray}
\Psi_a(x_a,t)= \tilde{\psi}_{a}(x_a,t) \exp (z_{a}(t)),
\label{mpnocoulomb}
\end{eqnarray}
where the term $z_{a}(t)$ is the (complex) purely time-dependent term that has no effect on the Bohmian trajectory $x_a[t]$, because this phase has no spatial dependence. Then, under the previous approximation, \eref{pseudo} can be simplified into the following equation for the computation of $\tilde{\psi}_{a}(x_a,t)$:
\begin{eqnarray}
i \hbar\frac{\partial \tilde{\psi}_{a}(x_a,t)}{\partial t}
= \left( -\frac{\hbar^2}{2m}\frac{\partial^2}{\partial {x^2_a}} + U_a (x_a, x_b[t]) \right) \tilde{\psi}_{a}(x_a,t),
\label{setpseudo}
\end{eqnarray}
Here, the potential energies can be $U_1 (x_1, x_2[t])=c(x_1-x_2[t])^2$ for $a=1$ and $U_2 (x_2, x_1[t])=c(x_1[t]-x_2)^2$ for $a=2$. The initial wave functions are ${\psi}_{1}(x_1,0)$  and ${\psi}_{2}(x_2,0)$ defined, both, form \eref{gausiana}. In particular, we consider $E_{o1}=0.06$ eV, $x_{o1}=50$ nm and $\sigma_{x1}=25$ nm for the first wave packet, and $E_{o2}=0.04$ eV, $x_{o2}=-50$ nm and $\sigma_{x2}=25$ nm for the second. In general, we need $N$-conditional wave functions to compute one $N$-particle Bohmian trajectory. If we change the initials positions, we need new $N$-conditional wave functions.

\begin{figure}[ht]%
\centering
\includegraphics*[width=8cm]{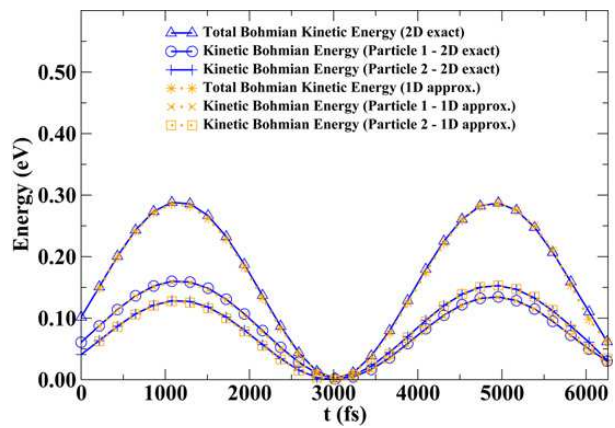}
\caption
{%
\footnotesize{(Color online) Time evolution of individual (ensemble averaged) Bohmian kinetic energies of identical two-electron system without exchange interaction under non-separable potential computed from $2D$ exact and $1D$ approximate solutions. }}
\label{fig12}
\end{figure}

In \fref{fig12}, we have plotted the same information than in \fref{fig11}  with our single-particle 1D approximation algorithm explained in \sref{review}. For this particular scenario, our simplest approximation for the unknown terms works perfectly and the agreement between 2D exact results and our 1D approximation is excellent. In general, potentials with small spatial variations are better adapted to the simplest $1D$ approximation of the term  $G_{a}$ and $J_{a}$ used in this work. We emphasize that the kinetic energy of the first and second particles are clearly distinguishable. We have compute the ensemble energies in order to justify that the algorithm is accurate not only for an arbitrarily selected set of Bohmian trajectory, but for most of them. In particular, the ensemble results are computed from $160000$ two-particle Bohmian trajectories.

\subsection{Algorithm to include exchange interaction in many-particle Bohmian trajectories}
\label{Algorithm}


Since \eref{pseudo} is valid for system with or without exchange interaction, one could look for  approximations to the terms $G_{a}$ and $J_{a}$ different from the simplest one mentioned above in order to incorporate the exchange interaction directly into \eref{pseudo}. However, from a computational point of view, such approximations seems quite difficult to implement. For example, in a system of fermions, we have seen that the wave function $\Psi_{a}(\vec{r}_a, t)$ becomes zero everywhere $\vec r_a$ is equal to the position of another trajectory. At these positions, because of their dependence on the inverse of the modulus, we would obtain $G_{a}\rightarrow \pm \infty$ and $J_{a}\rightarrow \pm \infty$. These infinities are difficult to treat numerically. In this subsection we present a different strategy that will be able to  capture the exchange
interaction avoiding the previous difficulties. The algorithm can be explained in four steps:

1.- The first step is developing an expression for $\Psi(\vec r_1,...\vec r_{N}, t)$ as a sum of wave functions. Each one of these wave function without symmetry. For example, let us define  $\Psi_{ns}(\vec r_1,...\vec r_{N}, t)$ as a many-particle wave function without any (bosonic or fermionic) symmetry. Then, we can construct a global wave function with exchange interaction using a sum of the term $\Psi_{ns}(\vec r_1,...\vec r_{N}, t)$ with all possible permutations of the positions:
\begin{eqnarray}
\label{wave3}
\Psi &=& C \sum\limits_{n=1}^{N!}  \Psi_{ns}(\vec{r}_{p(n)_1},..,\vec{r}_{p(n)_N},t) \;
 {\rm sign} \left( \vec{p}(n) \right),
\end{eqnarray}
Let us emphasize that  each term $\Psi_{ns}(\vec{r}_{p(n)_1},\vec{r}_{p(n)_2},...,\vec{r}_{p(n)_N},t)$ is also a solution of a many-particle  Schr\"{o}dinger equation for non-separable Hamiltonian without special exchange symmetry requirements. Finally, the conditional wave function $\Psi_{a}(\vec r_a, t)$ extracted from \eref{wave3} can be written as:

\begin{eqnarray}
\label{wave4}
\Psi_{a}(\vec r_a, t) = C \sum\limits_{n=1}^{N!} \Psi_{ns}(\vec{r}_{p(n)_1}[t],.,\vec{r}_{p(n)_e},.,\vec{r}_{p(n)_N}[t],t)
\times {\rm sign}\left( \vec{p}(n) \right).
\end{eqnarray}
We have substituted all positions by the corresponding trajectory except the degree of freedom $\vec r_{p(n)_e}=\vec r_a$.

2.- The second step is solving each wave function $\Psi_{ns}(\vec{r}_{p(n)_1}[t],...,\vec{r}_{p(n)_e},...,\vec{r}_{p(n)_N}[t],t)$ present in \eref{wave4} as a solution of \eref{pseudo}. Since $\Psi_{ns}$ has no exchange interaction, we can look for a solution similar to the one mentioned in the example in \sref{Moises_numerical_nonidentical}. Then, we can write $\Psi_{ns} \equiv \Psi_{ns}(\vec{r}_{p(n)_1}[t],...,\vec{r}_{p(n)_e},...,\vec{r}_{p(n)_N}[t],t)$ as:
\begin{eqnarray}
\Psi_{ns}= \tilde{\psi}_{p(n)_e,a}(\vec{r}_a,t) \exp (z_{p(n)_e,a}(t)),
\label{mpnocoulomb}
\end{eqnarray}
where the term $z_{p(n)_e,a}(t)$ is the (complex) purely time-dependent term related to $G_{a}^{''}(\vec{r}_{a}[t],\vec{X}_{a}[t],t)+J_{a}^{''}(\vec{r}_{a}[t],\vec{X}_{a}[t],t)$.  The subindex $a$ in $\tilde{\psi}_{p(n)_e,a}(\vec{r}_a,t)$ specifies which are the potential $U_a(\vec{r}_{a},\vec{X}_{a}[t],t)$ used when solving \eref{setpseudo}. The other subindex $p(n)_e$ identifies the initial wave function, as explained in next step.

3.- The third step is finding the initial wave function $\tilde{\psi}_{p(n)_e,a}(\vec{r}_a,0)$. When dealing with quantum transport,  we can assume that the initial many-particle wave function is located far from the active region (deep inside the reservoirs in a free space region) where it can be written as a Slater determinant (or permanent) as in \eref{slater} only during the time $t=0$. Then, we can easily realize that the initial state defining $\tilde{\psi}_{p(n)_e,a}(\vec{r}_a,0)\equiv \psi_{l}(\vec{r},0)$ is the particular wave packet of the ones defined in \eref{gausiana} which accomplishes ${p(n)_e}={a}$ (see Refs. \cite{llibrebohm} and \cite{note2}).

4.- The fourth step, once we know all wave functions $\tilde{\psi}_{p(n)_e,a}(\vec{r}_a,t)$,  is to compute the many-particle wave function $\Psi_{a}(\vec r_a, t)=\Psi(\vec r_1[t],...\vec r_{a-1}[t],\vec r_a,\vec r_{a+1}[t],..)$ in \eref{wave4} as:

\begin{eqnarray}
\label{prl_manywave}
\Psi_{a}(\vec{r}_a,\vec X_a[t],{t}) = C \sum_{n=1}^{N!}  \tilde{\psi}_{\vec{p}(n),a}(\vec{r}_a,t)
\times\exp { (z_{\vec{p}(n),a}(t)) } \; {\rm sign}\left( \vec{p}(n) \right).
\end{eqnarray}

We have to specify the values of the unknown phases $z_{p(n)_e,a}(t)$. We will fix these phases trying to satisfy the symmetry requirements of the Bohmian trajectories discussed in \sref{properties_trajectories}. In particular, we will demand that the observable results associated to different particles are indistinguishable. The following phases  $z_{\vec{p}(n),a}(t)$ accomplish the previous symmetry condition:
\begin{equation}
\label{prl_fases}
 \exp { (z_{\vec{p}(n),a}(t)) }=\prod^{N}_{k=1,k\neq a} \tilde \psi_{p(n)_e,a}(\vec{r}_k[t],t).
\end{equation}
In the \ref{appendix1} we show that this condition is enough to ensure that ensemble results of different particles are identical.

These are the four necessary steps needed to compute an $N-$particle Bohmian trajectory with exchange interaction for non-separable Hamiltonians. Let us discuss the number of conditional wave functions that we need for each $N-$particle Bohmian trajectory.  We realize that we have $N$ possible initial wave functions ${\psi}_{l}(\vec{r}_a,0)$ in \eref{prl_manywave}. Since the potential $U_{a}(\vec{r}_{a},\vec{X}_{a}[t],t)$ is invariant under the exchange of trajectories different than $\vec r_a[t]$, there are only $N$ different potentials needed.  Then, when computing the $N!$ functions $\tilde{\psi}_{p(n)_e,a}(\vec{r}_a,t)$ present from \eref{prl_manywave}, we realize that there are many repeated solutions. Therefore, there are $N \times N$ different wave functions $\tilde{\psi}_{l,a}(\vec{r}_a,t)$  that we have to solve in order to  compute \eref{prl_manywave}. The $N \times N$ correspond to $e=1,...,N$ different potentials and $a=1,....,N$ different initial wave packets.

In addition, it is important to notice what is the result of our algorithm when the non-separability of the Hamiltonian becomes negligible but the exchange interaction is still present. Then, we directly recover the Slater determinant (or permanent) defined in \eref{slater}. Finally, we want to emphasize that the algorithm for the inclusion of the exchange interaction is universal in the sense that exactly the same 4 steps have to be followed for any system.


\subsubsection{Test for non-separable harmonic potentials with exchange interaction}
\label{Moises_numerical}

\begin{figure}[ht]%
\centering
\includegraphics*[width=8cm]{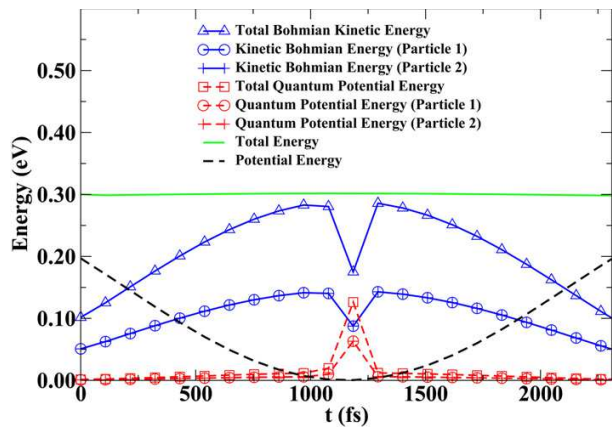}
\caption
{%
\footnotesize{(Color online) Time evolution of the total and individual (ensemble averaged) energies of identical two-electron system with exchange interaction under non-separable potential. } }
\label{fig13}
\end{figure}

In order to clarify the explanation of the exchange algorithm, we applied it to the same system discussed in \sref{Moises_numerical_nonidentical} but with the exchange interaction included.  In \fref{fig13}, we have plotted the ensemble results of the (Bohmian) kinetic energy, \eref{enkinetic}, the quantum  potential energy, \eref{enquantum}, computed directly from the two-particle wave function 2D exact solution of \eref{manyscho} for  two identical electrons (with parallel spins). In particular, we consider fermions with the some potential and initial wave packets that we discuss in  \sref{Moises_numerical_nonidentical}.  Now, the energies of particle 1 and 2 become indistinguishable.  In addition, we realize that the fact that Bohmian trajectories cannot cross the diagonal of the configuration space, implies a decrease/increase of the (Bohmian) kinetic/quantum energy when the wave function crosses the diagonal. This is the same effect discussed previously in \sref{Bohmian_trajectories_example}.

Finally, we compute the same results as in \fref{fig13} with the 1D approximated conditional wave functions discussed in \sref{Algorithm}. Now, we have to compute $4$ different functions, $\tilde{\psi}_{l,a}(x_a,0)$. The wave function $\tilde{\psi}_{1,1}(x_1,0)$ has initial wave function ${\psi}_{1}(x_1,0)$ defined from \eref{gausiana} and the potential $U_1 (x_1, x_2[t])=c(x_1-x_2[t])^2$. The wave function $\tilde{\psi}_{1,2}(x_2,t)$ has  the same initial state ${\psi}_{1}(x_2,0)$ but different potential energy $U_2 (x_2, x_1[t])=c(x_1[t]-x_2)^2$. Finally, $\tilde{\psi}_{\vec 2,1}(x_1,t)$ has initial state ${\psi}_{2}(x_1,0)$ and potential $U_2 (x_2, x_1[t])=c(x_1[t]-x_2)^2$, while $\tilde{\psi}_{\vec 2,2}(x_2,t)$ has the same initial state ${\psi}_{2}(x_2,0)$ and $U_1 (x_1, x_2[t])=c(x_1-x_2[t])^2$. The final wave functions $\Psi_{1}^l(x_1,t)$ and $\Psi_{2}^l(x_2,t)$ for the computation of the Bohmian trajectory, $x_1^l[t]$ and $x_2^l[t]$ are, respectively:

\begin{eqnarray}
\Psi_{1}^l(x_1,t)=C  \left(\tilde \psi_{1,1}^l(x_1,t)\tilde \psi_{2,2}^l(x_2[t],t)\
 -\tilde \psi_{2,1}^l(x_1,t)\tilde \psi_{1,2}^l(x_2[t],t) \right) , \\
\Psi_{2}^l(x_2,t)=C \left(\tilde \psi_{1,1}^l(x_1[t],t)\tilde \psi_{2,2}^l(x_2,t)
-\tilde \psi_{2,1}^l(x_1[t],t)\tilde \psi_{1,2}^l(x_2,t) \right).
\label{fin2}
\end{eqnarray}

\begin{figure}[ht]%
\centering
\includegraphics*[width=8cm]{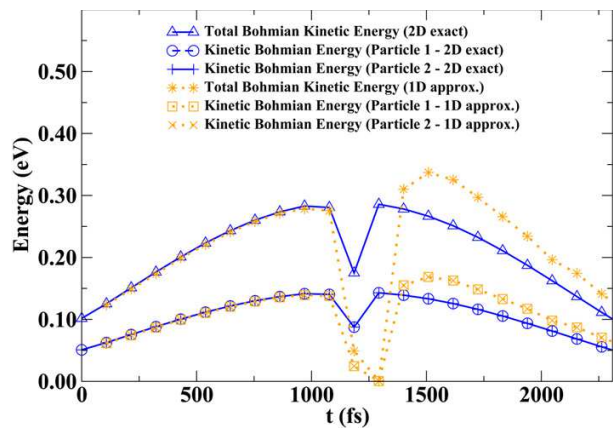}
\caption
{%
\footnotesize{(Color online) Time evolution of the individual (ensemble averaged) Bohmian kinetic energies of two-electron system with exchange interaction under non-separable potential computed from $2D$ exact and $1D$ approximate solutions.} }
\label{fig14}
\end{figure}

We emphasize that we require four single-particle wave functions for each 2-particle trajectory $\{x_1^l[t],x_2^l[t]\}$.  As discussed previously,  the algorithm with exchange scales as $N^2$. In \fref{fig14}, we have plotted the information about the energies for the same electrons discussed in \sref{Moises_numerical_nonidentical}. The agreement between the exact 2D results and the approximate 1D ones for identical particles is acceptable. Let us emphasize that the excellent agreement in \fref{fig12} and the results of \fref{fig14}, both, have been computed with the mentioned approximations on the terms  $G_{a}$ and $J_{a}$ in \eref{pseudo}. However, the algorithm with exchange interaction is more sensible to the approximations because we have to deal with  $\Psi_{a}^l(x_a,t)$ that are very close to zero at the diagonal of the configuration space. A small deviation (due to the approximate $G_{a}$ and $J_{a}$) in the value of the modulus close to zero becomes an amplified deviation in the velocity, as seen in \eref{Bohmianvelocity}, which is inversely proportional to the modulus. This difficulty is not present in the results of \fref{fig12} because, there, trajectories are not forced to be closer to regions where the modulus is zero. This difficulty is also manifested in the quantum potential. See the importance of the quantum potential in the $2D$ exact result with exchange interaction at $1100$ fs in \fref{fig13} (red lines), while the quantum potential is negligible in the $2D$ exact results without exchange interaction, as seen in \fref{fig11} (red lines).

All (ensemble) results presented in this work can be explained in terms of individual trajectories (each trajectory with different initial conditions). Thus, the error in the ensemble results in \fref{fig14} is due to errors in some individual trajectories, not all (mainly those trajectories starting outside of the center of the wave packet and arriving at regions where the wave function is almost zero). The total number of trajectories is 2$\times$160000 (computed from 4$\times$160000 conditional wave functions).  A wrong trajectory will never be converted into a correct one at a later time. On the contrary, one can expect that a correct trajectory at some particular time can become a wrong one at a later time due to the approximations in $G_{a}$ and $J_{a}$. In any case, a better approximation of the unknown terms $G_{a}$ and $J_{a}$ in \eref{pseudo} for systems without exchange will improve the accuracy of the algorithm with exchange. An interesting path to improve the approximations in the unknown terms $G_{a}$ and $J_{a}$ of the equation of the conditional wave function can be obtained by following Ref. \cite{travis}. It is showed there that the terms $G_{a}$ and $J_{a}$ in \eref{pseudo} can be computed, in principle, from a (infinite) set of coupled differential equations. A practical implementation will certainly require cutting the infinite set somewhere.

The most relevant feature is that the ensemble results for the first and second particle become indistinguishable with our 1D approximated conditional wave functions (see circle, plus, cross and square symbols in \fref{fig14}), although we perfectly distinguish Bohmian trajectories labeled as $\{x_1^l[t],x_2^l[t]\}$.

\section{Numerical results for electron transport simulation}
\label{alarcon_numerical_results}


Now, we use the general algorithm (with exchange interactions in non-separable Hamiltonians) to study quantum electron transport. The numerical implementation of the algorithm has been included into our simulator \textit{BITLLES} (Bohmian Interacting Transport for non-equiLibrium eLEctronic Structures) \cite{bitlles}. It is a general, versatile and time-dependent $3D$ electron transport simulator that allows the computation of DC, AC, transient and current and voltage fluctuations (noise) for nanoelectronic devices \cite{llibrebohm}. We compute the influence of the exchange interaction on the static and dynamic performance of a simple nano-resistor. We consider explicitly the Coulomb and exchange interaction among electrons inside the device active region. No other scattering mechanics is considered in the simulations. The details of the injection model for electrons (with a Binomial distribution) are explained in \ref{appendix3}.


\subsection{Definition of the simulation system with arbitrary spins}
\label{C6_alarcon_subsec_Computation of IV characteristic}

\begin{figure}[ht]%
\centering
\includegraphics*[width=8cm]{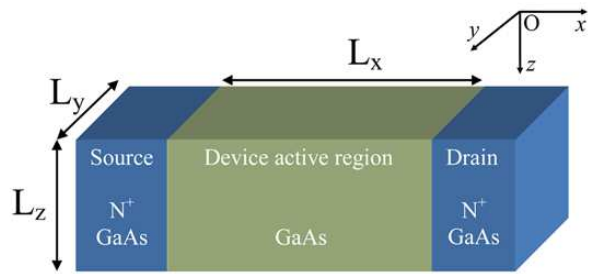}
\caption
{%
\footnotesize{(Color online) Scheme of a nano-resistor with  $N^{+}$ AsGa source/drain doped contacts and intrinsic AsGa in the device active region. The device dimension are $L_{x}$ = 30 nm,  $L_{y}=L_{z}$= 9 nm.  A Fermi level of $0.15 eV$ and room temperature are considered.} }
\label{alarcon_figure_nano_resistor}
\end{figure}

We consider a set of free electrons moving under the influence of the Coulomb and exchange interaction inside a device active region, when an external bias is applied between source and drain (see \fref{alarcon_figure_nano_resistor}). Electrons are injected into the device with an arbitrary spin. Therefore, in principle, we would have to consider several terms (each one with different spin distributions) in the many-particle wave function \eref{wave2} to treat properly the exchange interaction. The consideration of many terms in \eref{wave2} would imply an intractable computational burden, as mentioned in \ref{appendix2}. Alternatively, we can assume that the effect of the exchange interaction on the dynamics of an electron with spin up $\uparrow$ (down $\downarrow$) is due only to the other electrons with spin up $\uparrow$ (down $\downarrow$).  Therefore, in this work we will assume that the many-particle wave function can be separated into a product of spin-up and spin-down many-particle wave functions. For the example, we will consider that the wave function with arbitrary spins  $\Psi_{\uparrow_{1},\downarrow_{2}, \downarrow_{3},\uparrow_{4}}(\vec{r}_{1},\vec{r}_{2},\vec{r}_{3},\vec{r}_{4},t) \gamma(\uparrow_{1},\downarrow_{2},\downarrow_{3},\uparrow_{4})$, can be approximated as the product $\Psi_{\uparrow}(\vec{r}_{1},\vec r_4,t) \gamma(\uparrow_{1},\uparrow_{4})$ by $\Psi_{\downarrow}(\vec{r}_{2},\vec{r}_{3},t) \gamma(\downarrow_{2},\downarrow_{3})$:

\begin{eqnarray}
\Psi_{\uparrow_{1},\downarrow_{2}, \downarrow_{3},\uparrow_{4}}(\vec{r}_{1},\vec{r}_{2},\vec{r}_{3},\vec{r}_{4},t) \gamma(\uparrow_{1},\downarrow_{2},\downarrow_{3},\uparrow_{4}) \approx \nonumber\\
\Psi_{\uparrow}(\vec{r}_{1},\vec r_4,t) \gamma(\uparrow_{1},\uparrow_{4})
\Psi_{\downarrow}(\vec{r}_{2},\vec{r}_{3},t) \gamma(\downarrow_{2},\downarrow_{3}).
\label{2spin}
 \end{eqnarray}
From this approximation, we can apply our algorithm presented in \sref{Algorithm} to $\Psi_{\uparrow}$ and $\Psi_{\downarrow}$ independently. However, there are some terms in the left hand side of \eref{2spin} that are not present in the right hand side (see \ref{appendix2}). The approximation \eref{2spin} has already been tested in \cite{aalarcon09pps}. In the mentioned reference, it is shown that this approximation is almost an exact result when the normalized distance $d$ defined in \eref{distance} is larger than 1. For closer electrons, $d$ smaller than 1, an error appears in the computation of the Bohmian velocity \cite{aalarcon09pps}. On the other hand, we explicitly consider the Coulomb interaction among all the electrons, regardless of their spin. Therefore, we can expect that electrons try to be spatially separated, i.e. large $d$, because of the Coulomb repulsion among electrons. This argument provides an additional justification on the validity of \eref{2spin} when study quantum electron transport (with Coulomb and exchange interaction). Finally, let us notice that the number of particles varies as the simulation progress, i.e., we are dealing with a more complex many-particle wave function than that represented in \eref{wave2}.

In \fref{alarcon_figure_nano_resistor} we show a scheme of the nano-resistor that we will simulate with two GaAs source/drain doped contacts with a Fermi level of $0.15$ eV above the conduction band and a device active region of  intrinsic GaAs with length $L_{x}$ = 30 nm. Transport takes place in the $x$ direction with room temperature in all simulations. A single spherical band with $m^{*}_{GaAs}=0.067\;m_{0}$ (and $m_{0}$ the electron free mass) is considered. Because of the geometry $L_{x}\gg L_{y},L_{z}$, with $L_{y}=L_{z}$= 9 nm, energy confinement takes place in the lateral directions. We only take into account the first energy of the subband of GaAs with a value of $E_{1}=0.13$ eV just above the bottom of the conduction band.


\subsection{Computation of I-V characteristic}

\begin{figure}[ht]%
\centering
\includegraphics*[width=11cm]{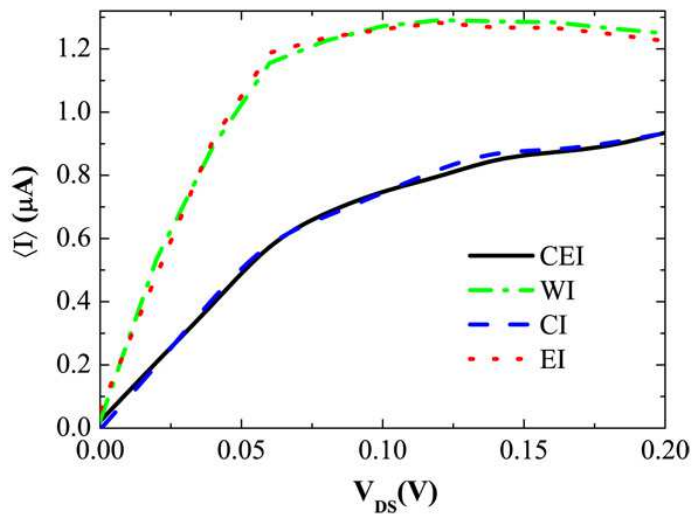}
\caption{%
 \footnotesize{(Color online) The average current as a function of the applied bias for the simulated system in four different situations: Coulomb and exchange interactions, without interactions, Coulomb interaction, and  exchange interaction.}}
\label{Fig_figure_exchange}
\end{figure}

In \fref{Fig_figure_exchange}, we present the time-averaged current $\left\langle I\right\rangle$ as a function of the external bias for four different scenarios:  Coulomb and exchange interactions (CEI), without exchange or Coulomb interactions (WI), with Coulomb interaction alone (CI) and with exchange interaction alone  (EI).
It is clear that the differences observed in the different curves are mainly a direct consequence of the Coulomb interaction. Somehow, for our particular device, the Coulomb interaction screens the effect of the exchange interaction because most of the electrons (not all) are already repelled by the Coulomb interaction. We emphasize that we are using Coulomb interaction beyond mean field, with self-interaction correction \cite{galbareda08PRB}.  The presence of Coulomb interaction tends to reduce the current because there are less electrons in the channel. Electrons repel each other.

\begin{figure}[ht]%
\centering
\includegraphics*[width=11cm]{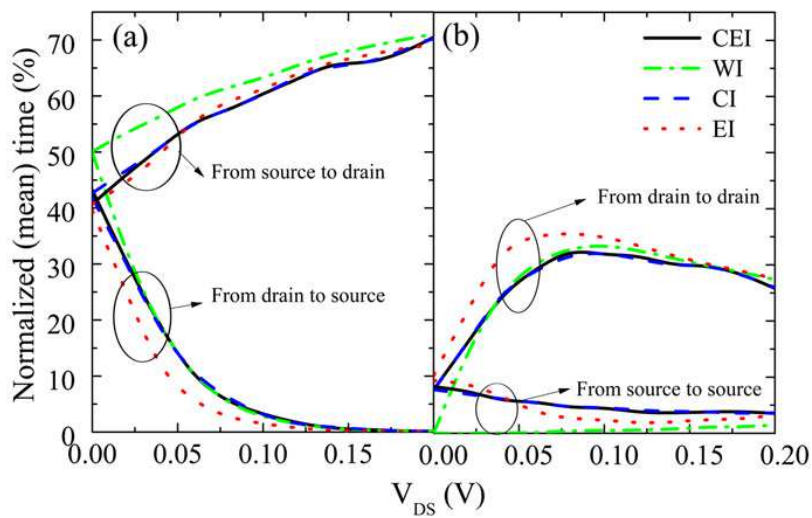}
\caption{%
 \footnotesize{(Color online) Normalized (mean) time spent by the electrons inside the simulating box as a function of the applied voltage for the four different scenarios discussed in \fref{Fig_figure_exchange}. (a) from drain to source, $d_{D/S}/d$, and from source to drain, $d_{S/D}/d$. (b) from the drain that have been finally bounced, $d_{D/D}/d$ and from the source that have been finally bounced, $d_{S/S}/d$}}
\label{Fig_figure_exchange_mean}
\end{figure}

For each simulated electron, we know when it enters the simulation box, $t^l_{in}$, and when it leaves $t^l_{out}$. The superindex $l$ indicates which Bohmian trajectory is associated to the electron.  Identically, we know whether the electron enters (leaves) the simulation box from the source (S) or drain (D) contacts. Then, we compute:
\begin{eqnarray}
d_{A/B}=\sum_{l_{A/B}}  \int \Theta(t-t_{in}^l)·\Theta(t_{out}^l-t) dt,
\label{particles}
\end{eqnarray}
where the Heaviside step function is  $\Theta(t)=1$ for positive times and $\Theta(t)=0$ for negatives ones. The time integral is over the whole simulation time. The sum is over all $l-$trajectories that have entered through the contact $A=\{S,D\}$ and leave through $B=\{S,D\}$.  In \fref{Fig_figure_exchange_mean}(a), we plot $d_{S/D}/d$ and $d_{D/S}/d$, and in \fref{Fig_figure_exchange_mean}(b) $d_{S/S}/d$ and $d_{D/D}/d$, where we have defined $d=d_{S/D}+d_{D/S}+d_{S/S}+d_{D/D}$. Electrons crossing the active region, $d_{S/D}$ and $d_{D/S}$, are responsible for the DC (i.e. zero frequency) behavior. On the other hand, electrons that do not cross the device active region, $d_{S/S}$ and $d_{D/D}$, do not contribute to DC, but only to high-frequency dynamics.

At zero bias, as seen in \fref{Fig_figure_exchange_mean}(a), without interaction (WI), half of electrons are transmitted from source to drain and half from drain to source. No reflected electrons. This is not true for the rest of scenarios. The EI simulation provides reflected electrons because of the effect seen in \sref{Bohmian_trajectories_example}, that forbids electrons from occupying the same positions (i.e. the diagonals points of the configuration space) and some of the electrons are finally bounced. Let us emphasize that the mean number of electrons in the active region of the quantum wire of \fref{alarcon_figure_nano_resistor} can be very small. From the current in \fref{Fig_figure_exchange}, we can compute the rate of transmitted electrons which is quite similar to their transit time. The CI (and CEI) simulation shows reflected because of the Coulomb repulsion among electrons. Finally, we focus on set of plots $d_{D/D}/d$ (from drain to drain) in \fref{Fig_figure_exchange_mean}(b). At a 0.05 V, the EI simulation has a larger value $d_{D/D}/d$ than the others. This result does not implies a larger number of reflected particles with EI simulation, but only that these EI reflected particles spent more time inside the active region than, for example, the CI and CEI ones. The reflection in the EI simulation occurs when the particles are really very close, see \fref{Fig6}(a), while the reflection in the CI simulation occurs for particles with a larger spatial separation. The Coulomb interaction has a longer range than the particular (exchange interaction) effect seen in \fref{Fig6}(a). Thus, these CI and CEI particles spent less time in the active region. This result justifies again why the Coulomb interaction, somehow, screens the possible effects of the exchange interaction.


\subsection{Computation of the noise}


\begin{figure*}[ht]%
\begin{minipage}{15.8cm}
\centering
\includegraphics*[width=7cm]{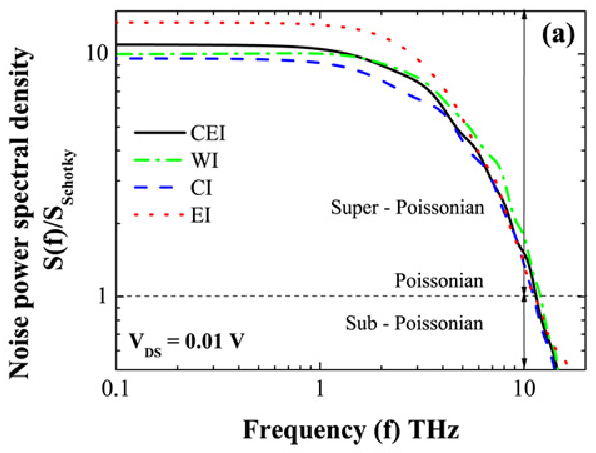}
\includegraphics*[width=7cm]{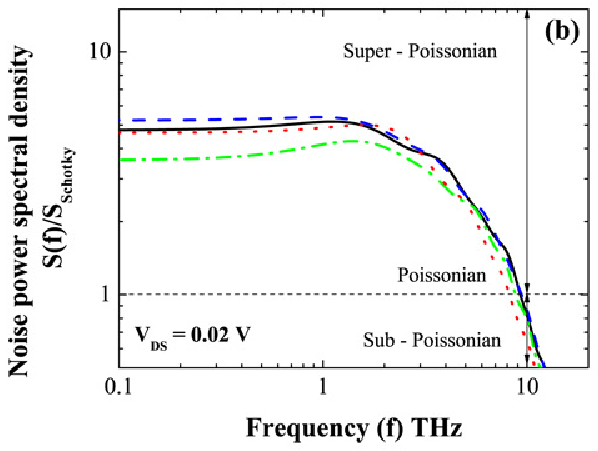}
\includegraphics*[width=7cm]{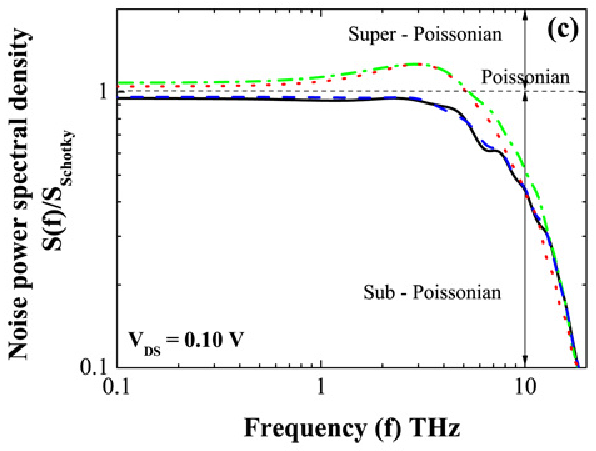}
\includegraphics*[width=7cm]{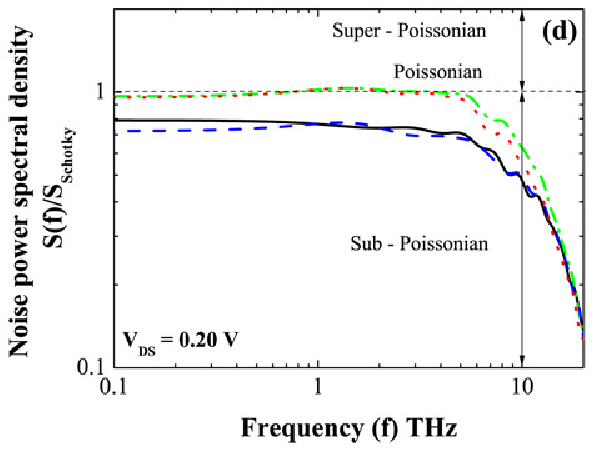}
\caption{%
  \footnotesize{(Color online) Noise power spectral density as a function of the frequency for four applied bias. (a) $V_{DS}$ = 0.01 V (b) $V_{DS}$ = 0.02 V (c) $V_{DS}$ = 0.1 V (d) and $V_{DS}$ = 0.2 V.}}
\label{Fig_figure_exchange_1}
\end{minipage}
\end{figure*}

From the time-dependent current $I(t)$ provided by the Monte Carlo \textit{BITLLES} simulator we can also study the noise characteristics of this simple nano-resistor in a very simple way \cite{comment2}. The fluctuations of the current can be easily obtained from the autocorrelation function $R(\tau)=\overline{\Delta I(t)\Delta I(t+\tau)}$ of the current fluctuations $\Delta I(t)=I(t)-\left\langle I \right\rangle$ being $I(t)$ the instantaneous current provided by the quantum Monte Carlo BITLLES simulator and $\left\langle I\right\rangle$ the time-average DC current computed above.
The Fourier transform of this autocorrelation function $R(\tau)$ is the noise power spectral density of $S(f)=\int_{-\infty}^\infty R(\tau)e^{-i 2 \pi f \tau} \ d\tau$.
Finally, the Fano factor, defined as the ratio $\gamma=S(0)/S_{schottky}$ with $S_{Schottky}(f)=2q\left\langle I\right\rangle$, can be computed.

In \fref{Fig_figure_exchange_1} we show the noise power spectral density as a function of the frequency for four different bias.
The Fano factor decreases as we increase the voltage. Since we are simulating at room temperature, for low bias tending to zero, the  current tends to zero while the thermal noise is still present giving a Fano factor tending to infinite. For high bias, the fluctuations follows the Binomial distribution \eref{eq::Fermi1} that roughly  tends to a Poissonian one for electrons with energy above the Fermi level  (as is the case for the sample with confinement discussed here).
Deviations from this behavior are due to correlations originated from Coulomb and exchange interactions. First, we observe in \fref{Fig_figure_exchange_1} that, similarly to DC, the the most important effect on the Fano factor is due to the Coulomb interaction (see CI and CEI in \fref{Fig_figure_exchange_1}). As we see in \fref{Fig_figure_exchange_1}(d), the Fano factor is lower than one with CEI or CI simulations. It is well-known that the Coulomb repulsion between electrons tends to space them more regularly rather than strictly at random, and to evidence a sub-Poissonian statistics \cite{nazarov,tomas}. However, we see that the small  deviations between the CEI and CI results are influenced by the exchange interaction present in the simulation box. When EI is larger than CI, the CEI is larger than CI, and vice-verse. Secondly, we observe that the effect of the exchange interaction in the noise is more important at low bias. This result has the same explanation that we explain in the previous \fref{Fig_figure_exchange} and \fref{Fig_figure_exchange_mean}. Thirdly, in \fref{Fig_figure_exchange_1}(b) and (c), at intermediate bias, we see a peak of the noise power spectral density above $1$THz. One can realize that this peak in the fluctuations appears when only WI or EI are considered. The origin of this peak is the increment of the number of reflected electrons because they cannot occupy identical positions. These bounced electrons do not affect the zero (low) frequency fluctuations, but they affect the high-frequency values. This increment of reflected particles can be seen in \fref{Fig_figure_exchange_mean}(b).

Finally, as discussed in the injection process in \ref{appendix3}, we want to mention that wave packets with identical central positions and wave vectors are injected with a temporal separation between them equal to $t_0$ that tries to avoids relevant exchange interaction effects among them inside the simulating box.


\section{Discussion and conclusions}
\label{Conclusions}


In this work we have presented an algorithm for introducing the exchange interaction into the many-particle quantum (Bohmian) trajectories with an universal protocol valid for any quantum system, with separable or non-separable Hamiltonians, for either fermions or bosons. In principle, one could thought in introducing the exchange interaction into the terms $G_{a}$ and $J_{a}$ present in \eref{pseudo} that define the conditional wave function,  in a similar way as the  exchange-correlation functional is introduced into the Hamiltonian of a single-particle pseudo-Schr\"{o}dinger equation used by DFT. However, There is no clear prescription on how to define directly the terms $G_{a}$ and $J_{a}$  with exchange interaction. Alternatively, in this work we follow a different path. We compute $N \times N$ conditional wave function solutions of \eref{pseudo} [with a very simple approximation for $G_{a}$ and $J_{a}$, leading to \eref{setpseudo}] without symmetry requirements. However,  the global conditional wave function constructed as a combination of them,  \eref{prl_manywave}, can satisfy the exchange requirements, after a proper guess for some phases. We have shown that the phases defined from \eref{prl_fases} satisfy, by construction, three very satisfactory properties. First, in the case of separable Hamiltonians, it directly leads to the standard Slater determinants for computing many-particle wave functions (or the permanent for bosons). Second, for fermions with non-separable Hamiltonians, it guarantees that the probability presence of the many-particle wave function in the diagonal points of the configuration space is zero.  Third, in any scenario, the Bohmian trajectories satisfies the expected symmetry property when initial positions are interchanged. This last property ensures that observable results computed from Bohmian trajectories are indistinguishable, as seen in \fref{fig14}. Note that the selection of the phases in \eref{prl_fases} is not unique. In the \ref{appendix1}, we have shown another possibility, but it does not satisfy the second property discussed above.

An improvement on the simple approximation used to compute $G_{a}$ and $J_{a}$ when constructing the conditional (Bohmian) wave functions without exchange interaction would also improve the accuracy of the algorithm with exchange presented here.

As a practical demonstration of the numerical viability of the algorithm discussed here for quantum transport (in a far-from equilibrium open system),  the current and its fluctuations are computed for a nano-resistor, with exchange and Coulomb interactions. For this simple device, the effects of the exchange interaction are mainly screened by the Coulomb interaction. In any case, the main conclusions that one can extract from these numerical results, applied to this very simple device, is that the algorithm explained in this work can be perfectly implemented for a number of electrons on the order of 20-30.  This requires solving $30 \times 30 \approx 1000$ single particle conditional wave functions, which can still be handled with normal computing facilities and simulation times on the order of few hours for each simulation bias.

\section*{acknowledgement}
The present work have greatly beneficed from discussions with W. Struyve and G. Albareda.  This work was supported through Spanish project MICINN TEC2012-31330.

\appendix
\section{Indistinguishable results computed from the approximate conditional Bohmian wave functions using \eref{prl_fases}}
\label{appendix1}

The requirement for ensuring that ensemble results computed from Bohmian trajectories are indistinguishable is that the $N$-particle Bohmian trajectories are symmetric under the interchange of their initial positions:  We consider two different $N$-particle Bohmian trajectories, whose initial positions are $\vec X^l[0]$ and $\vec X^f[0]$. In particular, we consider $\vec r_k^l[0]=\vec r_k^f[0]$ for all $k$ except $\vec r_j^l[0]=\vec r_h^f[0]$ and $\vec r_h^l[0]=\vec r_j^f[0]$. Then, the sufficient condition to ensure that observable results of different particles are indistinguishable is ensuring that $\vec r_k^l[t]=\vec r_k^f[t]=\vec r_k[t]$ for all $k$ except $\vec r_j^l[t]=\vec r_h^f[t]$ and $\vec r_h^l[t]=\vec r_j^f[t]$. In this appendix, we show that this last property is guaranteed by our proposal of using \eref{prl_fases} for fixing the unknown phases $z_{p(n)_e,a}(t)$.

At the initial time, by construction (see step 3 in \sref{Algorithm}), it is obvious that \eref{prl_fases} provides Bohmian trajectories with the desired property. Let us demonstrate that this relation between trajectories at the initial time is also true at a later time. We assume that the Bohmian trajectories have the desired property at the time $t'$ (for example, $t'=0$). Then, we want to demonstrate that $\vec r_j^l[t]=\vec r_h^f[t]$ and $\vec r_h^l[t]=\vec r_j^f[t]$ at a later time $t=t'+dt$. In fact, we only have to demonstrate that the velocities of the particles $j$ and $h$ at $t'$ are interchanged. According to \eref{prl_fases}, the conditional wave function used to compute the velocity $\vec v_j^l(\vec X,t')|_{\vec X=\vec X^l[t']}$ with the initial positions $\vec X^l[t']=\{\vec r_1[t'],...,\vec r_{\alpha}[t'],...,\vec r_{\beta}[t]...\}$ is:

\begin{eqnarray}
\Psi_{j}^l(\vec r_j,t\rq{}) &=& C \sum\limits_{n=1}^{N!}\tilde \psi_{ p(n)_e,\alpha}(\vec r_1[t],t\rq{}),.., \tilde \psi_{ p(n)_e,\alpha}(\vec r_j,t\rq{}) \nonumber \\
&,..,&\tilde \psi_{ p(n)_e,\alpha}(\vec r_{\beta}[t],t\rq{})..,\tilde \psi_{p(n)_e,\alpha}(\vec r_N[t],t\rq{})
\times{\rm sign}\left( \vec{p}(n) \right),
\label{we1}
\end{eqnarray}
where we have defined $\vec r_j^l[t]=\vec r_h^f[t] \equiv \vec r_{\alpha}[t]$ and $\vec r_h^l[t]=\vec r_j^f[t]\equiv \vec r_{\beta}[t]$. Identically, we have $\vec v_h^l(\vec X,t')|_{\vec X=\vec X^l[t']}$:

\begin{eqnarray}
\Psi_{h}^l(\vec r_h,t\rq{}) &=& C \sum\limits_{n=1}^{N!}\tilde \psi_{ p(n)_e,\beta}(\vec r_1[t],t\rq{}),.., \tilde \psi_{ p(n)_e,\beta}(\vec r_{\alpha}[t],t\rq{}) \nonumber \\
&,..,&\tilde \psi_{ p(n)_e,\beta}(\vec r_h,t\rq{})..,\tilde \psi_{p(n)_e,\beta}(\vec r_N[t],t\rq{})\times{\rm sign}\left( \vec{p}(n) \right).
\label{we2}
\end{eqnarray}

On the other hand, for the many-particle Bohmian trajectory with initial conditions $\vec X^f[0]=\{\vec r_1[0],...,\vec r_{\beta}[0],...,\vec r_{\alpha}[0]...\}$,  we have for $\vec v_j^f(\vec X,t')|_{\vec X=\vec X^f[t']}$:

\begin{eqnarray}
\Psi_{j}^f(\vec r_j,t\rq{}) &=& C \sum\limits_{n=1}^{N!}\tilde \psi_{ p(n)_e,\beta}(\vec r_1[t],t\rq{}),.., \tilde \psi_{ p(n)_e,\beta}(\vec r_j,t\rq{}) \nonumber \\
&,..,&\tilde \psi_{ p(n)_e,\beta}(\vec r_{\alpha}[t],t\rq{})..,\tilde \psi_{p(n)_e,\beta}(\vec r_N[t],t\rq{})\times {\rm sign}\left( \vec{p}(n) \right).
 \label{we3}
\end{eqnarray}
and for $\vec v_h^f(\vec X,t')|_{\vec X=\vec X^f[t']}$:

\begin{eqnarray}
\Psi_{h}^f(\vec r_h,t\rq{}) &=& C \sum\limits_{n=1}^{N!}\tilde \psi_{ p(n)_e,\alpha}(\vec r_1[t],t\rq{}),.., \tilde \psi_{ p(n)_e,\alpha}(\vec r_{\beta}[t],t\rq{}) \nonumber \\
&,..,&\tilde \psi_{ p(n)_e,\alpha}(\vec r_h,t\rq{})..,\tilde \psi_{p(n)_e,\alpha}(\vec r_N[t],t\rq{})\times{\rm sign}\left( \vec{p}(n) \right).
\label{we4}
\end{eqnarray}
As expected, it becomes obvious that $\vec v_j^l(\vec r_j,t')=\vec v_h^f(\vec r_h,t')$ and $\vec v_h^l(\vec r_h,t')=\vec v_j^f(\vec r_j,t')$. Q.E.D.

The use of expression \eref{prl_fases} does also provide the additional property that $\Psi_{j}^l(\vec r_j,t)$ becomes zero whenever $\vec r_j=\vec r_k[t]$ for $j \neq k$.  It can also be shown that  $\exp { (z_{\vec{p}(n),a}(t)) }=\prod^{N}_{k=1,k\neq a} \tilde \psi_{p(n)_e,k}(\vec{r}_k[t],t)$ does also provide the required symmetry condition for the trajectories, but it does not guarantee that the conditional wave function is zero for fermions at the coincident points. Thus, \eref{prl_fases} is preferred. Let us emphasize that ensuring that observable results become indistinguishable is not enough to ensure that our algorithm provides the correct results. Apart from the requirement mentioned here, Bohmian trajectories have to satisfy other properties as for example their non-crossing property discussed in \sref{properties_trajectories}.

\section{Exchange interaction for electrons with different spin}
\label{appendix2}

Even for systems without spin-orbit interaction and when we are not interested in the time evaluation of the spins, we cannot neglect the spin degrees of freedom of the electrons (with arbitrary spins) because the symmetry of the overall wave function depends on the exchange properties of the orbital part and spin component. To understand the complexity of computing the antisymmetry wave function with spins in different directions, we present an example for three electrons, one with spin up ($\uparrow_{j}$) and the others two with spin down ($\downarrow_{j}$). Then, the global antisymmetric wave function in \eref{wave2}, for this particular case, can be written as:
\begin{eqnarray}
\label{B1}
 \Phi (x_1 ,x_2 ,x_3;\uparrow_1 , \downarrow_2 , \downarrow_3 )=\nonumber\\+ \psi _1 (x_1 )\psi _2 (x_2 )\psi _3 (x_3 )\gamma( \uparrow_1,  \downarrow_2,  \downarrow_3 )   - \psi _1 (x_1 )\psi _2 (x_3 )\psi _3 (x_2)\gamma ( \uparrow_1, \downarrow_3,  \downarrow_2)\nonumber   \\
  - \psi _1 (x_2 )\psi _2 (x_1 )\psi _3 (x_3 ) \gamma( \downarrow_2,  \uparrow_1,  \downarrow_3 )   + \psi _1 (x_3 )\psi _2 (x_1 )\psi _3 (x2 ) \gamma ( \downarrow_3,  \uparrow_1,  \downarrow_2 ) \\
  + \psi _1 (x_2 )\psi _2 (x_3 )\psi _3 (x_1 ) \gamma ( \downarrow_2,  \downarrow_3,  \uparrow_1 )   - \psi _1 (x_3 )\psi _2 (x_2 )\psi _3 (x_1 ) \gamma( \downarrow_3,  \downarrow_2,  \uparrow_1 ). \nonumber
 \end{eqnarray}
We define the orbital wave functions $\psi _l (x_l )$ as the Gaussian wave packets of \eref{gausiana}. \Eref{B1} has $3!$ terms, each one composed of the product of an orbital function by a spin function.
Next, we compute the total norm taking into account the $3!3!$ products of permutations.
For this purpose we have to multiply the orbital parts and the spin parts separately.
Due to orthogonality, the product of the spin part can be either $0$ and $1$.
The final result is:
\begin{eqnarray}
 \left| {\Phi (x_1 ,x_2 ,x_3;\uparrow_1 , \downarrow_2 , \downarrow_3 )} \right|^2  =\nonumber  \\
  + \left[ {\psi _1 ^* (x_1 )\psi _2 ^* (x_2 )\psi _3 ^* (x_3 )} \right]\psi _1 (x_1 )\psi _2 (x_2 )\psi _3 (x_3 )\nonumber \\
  - \left[ {\psi _1 ^* (x_1 )\psi _2 ^* (x_2 )\psi _3 ^* (x_3 )} \right]\psi _1 (x_1 )\psi _2 (x_3 )\psi _3 (x_2 )\nonumber \\
  - \left[ {\psi _1 ^* (x_1 )\psi _2 ^* (x_3 )\psi _3 ^* (x_2 )} \right]\psi _1 (x_1 )\psi _2 (x_2 )\psi _3 (x_3 )\nonumber \\
  + \left[ {\psi _1 ^* (x_1 )\psi _2 ^* (x_3 )\psi _3 ^* (x_2 )} \right]\psi _1 (x_1 )\psi _2 (x_3 )\psi _3 (x_2 )\nonumber \\
  + \left[ {\psi _1 ^* (x_2 )\psi _2 ^* (x_1 )\psi _3 ^* (x_3 )} \right]\psi _1 (x_2 )\psi _2 (x_1 )\psi _3 (x_3 ) \\
  - \left[ {\psi _1 ^* (x_2 )\psi _2 ^* (x_1 )\psi _3 ^* (x_3 )} \right]\psi _1 (x_3 )\psi _2 (x_1 )\psi _3 (x_2 )\nonumber \\
  - \left[ {\psi _1 ^* (x_3 )\psi _2 ^* (x_1 )\psi _3 ^* (x_2 )} \right]\psi _1 (x_2 )\psi _2 (x_1 )\psi _3 (x_3 )\nonumber \\
  + \left[ {\psi _1 ^* (x_3 )\psi _2 ^* (x_1 )\psi _3 ^* (x_2 )} \right]\psi _1 (x_3 )\psi _2 (x_1 )\psi _3 (x_2 )\nonumber \\
  + \left[ {\psi _1 ^* (x_2 )\psi _2 ^* (x_3 )\psi _3 ^* (x_1 )} \right]\psi _1 (x_2 )\psi _2 (x_3 )\psi _3 (x_1 )\nonumber \\
  - \left[ {\psi _1 ^* (x_2 )\psi _2 ^* (x_3 )\psi _3 ^* (x_1 )} \right]\psi _1 (x_3 )\psi _2 (x_2 )\psi _3 (x_1 )\nonumber \\
  - \left[ {\psi _1 ^* (x_3 )\psi _2 ^* (x_2 )\psi _3 ^* (x_1 )} \right]\psi _1 (x_2 )\psi _2 (x_3 )\psi _3 (x_1 )\nonumber \\
  + \left[ {\psi _1 ^* (x_3 )\psi _2 ^* (x_2 )\psi _3 ^* (x_1 )} \right]\psi _1 (x_3 )\psi _2 (x_2 )\psi _3 (x_1 )\nonumber.
  \label{alarcon_modulo}
 \end{eqnarray}
Let us notice that, in principle, we had to keep  $3!3!=6^2=36$ terms.
However, only these terms whose product of spin parts is $1$ are present in \eref{alarcon_modulo}. The evaluation of the product of the spin parts have to be done explicitly, term by term, with no possibility of simplification. For example, the product of $\gamma( \uparrow_1,  \downarrow_2,  \downarrow_3 )$ by $\gamma ( \uparrow_1, \downarrow_3,  \downarrow_2)$ is $1$, while the product of $\gamma( \uparrow_1,  \downarrow_2,  \downarrow_3 )$ by $\gamma( \downarrow_2,  \uparrow_1,  \downarrow_3 )$ is $0$. However, if we increase the number of electrons, the practical computation of the previous expression is computationally inaccessible. Note that $N=8$ gives $8!^{2} = 40320^{2}$ terms.

In \eref{2spin} we  provide a (computationally accessible) approximation to treat wave functions with spin of different orientations.  For example, if we assume that there is no exchange interaction between spin up and spin down components, then:
\begin{eqnarray}
 \bar \Phi(x_1 ,x_2 ,x_3;\uparrow_1 , \downarrow_2 , \downarrow_3 )=\nonumber\\+ \psi _1 (x_1 )\gamma( \uparrow_1) \left(\psi _2 (x_2 )\psi _3 (x_3 )\gamma(\downarrow_2,  \downarrow_3 )   - \psi _2 (x_3 )\psi _3 (x_2)\gamma (\downarrow_3,  \downarrow_2) \right).
 \end{eqnarray}
Now, the norm would be:
\begin{eqnarray}
 \left| {\bar \Phi (x_1 ,x_2 ,x_3;\uparrow_1 , \downarrow_2 , \downarrow_3 )} \right|^2  =\nonumber  \\
  + \left[ {\psi _1 ^* (x_1 )\psi _2 ^* (x_2 )\psi _3 ^* (x_3 )} \right]\psi _1 (x_1 )\psi _2 (x_2 )\psi _3 (x_3 )\nonumber \\
  - \left[ {\psi _1 ^* (x_1 )\psi _2 ^* (x_2 )\psi _3 ^* (x_3 )} \right]\psi _1 (x_1 )\psi _2 (x_3 )\psi _3 (x_2 )\nonumber \\
  - \left[ {\psi _1 ^* (x_1 )\psi _2 ^* (x_3 )\psi _3 ^* (x_2 )} \right]\psi _1 (x_1 )\psi _2 (x_2 )\psi _3 (x_3 )\nonumber \\
  + \left[ {\psi _1 ^* (x_1 )\psi _2 ^* (x_3 )\psi _3 ^* (x_2 )} \right]\psi _1 (x_1 )\psi _2 (x_3 )\psi _3 (x_2 )\nonumber \\
 \label{alarcon_modulo1}
 \end{eqnarray}
where there are terms on \eref{alarcon_modulo} that are not present in \eref{alarcon_modulo1}.  However, in Ref. \cite{aalarcon09pps} we show that the   approximation \eref{2spin} provides a quite reasonable approximation for the computation of the Bohmian velocity. Expression (\ref{alarcon_modulo}) keep the most relevant terms of the exchange interaction. \\

\section{Electron injection probability}
\label{appendix3}

Our Bohmian algorithm requires each electron to be described by a (conditional) wave function plus a Bohmian trajectory. Every time an electron with a particular initial wave function is selected to enter the device active region, an initial position for the Bohmian trajectory associated to this wave packet has to be randomly selected according to  'quantum equilibrium hypothesis' \cite{HollandPR1993,goldstein} discussed in \sref{Many-particle trajectories}. Accordingly, this initial position is more frequently found to be around the center of the wave packet than in the borders. Next, we explain how the wave packets are selected.

The (central) kinetic energy $E_o$ of the wave packet is related to the (central) wave vectors $k_{o}$ by $E_o=(\hbar k_{o})^2/(2 m^*)$ with $m^*$ the effective mass. For each contact, we select a flat potential region in an (non-physical) extension of the simulation box  (for $x<0$ in the source and $x>L_x$ in the drain of \fref{alarcon_figure_nano_resistor}). Let us notice that in a flat potential region without interaction, a single-particle wave packets is exactly the normalized conditional (Bohmian) wave function discussed in this work. The initial Gaussian wave packet is defined in this flat potential region (deep inside the contact) following the analytical expression (\ref{gausiana}). In particular, the central position of all wave packets is selected $x_o=100\;nm$ far from the border of the active region (inside the contacts) and the spatial dispersion of the wave packet is $\sigma_x=25\; nm$ (i.e. the wave packet is somehow similar to a scattering state). The only two additional parameters that we still have to fix to fully define the wave packet are the central kinetic energy $E_o$ of the wave packet and the injecting time when the electron \emph{effectively} enters the simulation box. The selection of the energy $E_o$ has to satisfy the Fermi-Dirac occupation function $f (E_o)$ that depends on the (quasi) Fermi energy and temperature. The selection of the time when the electron is injected is a a bit more complex.

Let us define $t_0$ as the minimum temporal separation between the injection of two wave packets whose central wave vectors and central positions fit into the following particular phase-space cell $k_o  \in [k_b ,\,\,k_b  + \Delta k )$ and $x_c \in [x_b ,\,x_b  + \Delta x)$, being $x_b$ the left border of the simulation region. For a 1D system, the value of $t_0$ can be easily estimated. The number of electrons $n_{1D}$ in the particular phase space cell $\Delta k \cdot\Delta x$ is $n_{1D}  = 2\cdot\Delta k \cdot\Delta x/(2\pi )$ where the factor  2 takes into account the spin degeneracy. These electrons have been injected into $\Delta x$ during the time interval $\Delta t$ defined as the time needed for electrons with velocity $v_x  = \Delta x/\Delta t = \hbar \,k_o /m$ to travel a distance $\Delta x$. Therefore, the minimum temporal separation, $t_0$, between the injection of two electrons into the previous cell is $\Delta t$ divided by the maximum number $n_{1D}$ of electrons:
\begin{equation}
\label{eq::Fermi2}
 {t_0}  = \frac{{\Delta t}}{{n_{{\rm{1D}}} }} = \left( {\frac{1}{\pi }\frac{{\hbar \,k_o }}{{m }}\,\,\Delta k } \right)^{ - 1}.
\end{equation}
It is very instructive to understand the minimum temporal separation $t_0$ in (\ref{eq::Fermi2}) as a consequence of the wave packet version of the Pauli principle discussed in \sref{total energy}.  The simultaneous injection of two electrons with similar central positions and central momentums would require such a huge amount of energy that its probability is almost zero (see  \fref{Fig_ec_d}). In other words, a subsequent electron with central position and central momentum equal to the preceding ones can only be injected after a time interval given by $t_0$.

 The injection of electrons (from the mentioned phase-space cell) at multiple times of $t_o$ depends finally on the statistics imposed by the Fermi-Dirac function mentioned above. During each attempt of injection at multiples of $t_o$, we select a random number $r$, and the electron is effectively injected only if $f(E_o)>r$. The mathematical definition of the  rate and randomness of the injection process are given by the following binomial probability $P(E_o,N_{\tau},\tau)$ (See Ref. \cite{xoriols07SEE}):
\begin{equation}\label{eq::Fermi1}
P(E_o,N_{\tau},\tau ) = \frac{{M_\tau  !}}{{N_{\tau}!\cdot(M_\tau   - N_{\tau})!}}f(E_o)^{N_{\tau}} \left( {1 - f(E_o)} \right)^{M_\tau   - N_{\tau}}.
\end{equation}
This expression defines the probability that $N_{\tau}$ electrons (from the mentioned phase-space cell) are \emph{effectively} injected into the active region during the time interval $\tau$. The parameter $M_\tau$ is the number of attempts of injecting electrons during this time interval $\tau$, defined as a natural number that rounds the quotient $\tau /t_o$ to the nearest natural number towards zero. The number of injected electrons can be $N_{\tau} = 1,2,.... \le M_\tau$.  More details can be found in Ref. \cite{xoriols07SEE}. Finally, let us clarify that \eref{eq::Fermi1} does only specify the injecting probability. The transmission probability with a certain energy depends on the injecting probability and also on all complex (Coulomb and exchange) phenomena explained along the text.


\begin{thebibliography}{99}
\bibitem{comment}
In this work, the quantum trajectories are only used to reproduce ensemble results without any ontological implication. Then, the adjective Bohmian and hydrodynamic are fully interchangeable in this work. However, we use the first one only because the concept of conditional wave function, which is fundamental in this work, is mainly developed for Bohmian mechanics.
\bibitem{notespin}
We use the basis of individual spins instead of the basis of the total spin because we are interested in open systems, where particles (spins) are interchanged with reservoirs. Then, the requirement of constant total spin (implicit in the total spin basis) is not at all evident.
\bibitem{note2}
For example, the permutation $\vec p(n)={\{3,2,1\}}$ on the conditional wave function used to compute the trajectory $\vec r_1[t]$ has the initial wave packet $\psi_{3}(\vec{r}_1,t)$ because ${p(n)_3}={1}$.
\bibitem{comment2}
By construction, an ensemble of Bohmian trajectories reproduces the probability presence of the many-particle wave function at any time and also the correlations. The only approximation in our computations of the current is that the (continuous) time-dependent measurement of the current provides a negligible back-action on the simulated system \cite{llibrebohm}.
\end{thebibliography}
\end{document}